\documentclass[aps,twocolumn,nofootinbib,showpacs,floatfix]{revtex4}

\usepackage[centertags]{amsmath}
\usepackage{amsfonts}
\usepackage{graphicx}
\usepackage[hypertex]{hyperref}
\usepackage{psfrag}

\newcommand{\bea}{\begin{eqnarray}}
\newcommand{\eea}{\end{eqnarray}}
\newcommand{\be}{\begin{equation}}
\newcommand{\ee}{\end{equation}}

\newcommand{\Lo}{\Lambda_{0}}

\def\spose#1{\hbox to 0pt{#1\hss}}
\def\ltapprox{\lesssim}

\newcommand{\ve}[1]{\ensuremath{\mathbf{#1}}}
\newcommand{\D}[1][ ]{\ensuremath{\mathrm{d}^{#1} }}
\newcommand{\tdyn}{\ensuremath{\tau_\text{dyn}}}

\begin{document}

\title{Gravitational Dynamics of an Infinite Shuffled Lattice of Particles}

\author{T. Baertschiger}
\affiliation{Dipartimento di Fisica, Universit\`a ``La Sapienza'',
P.le A. Moro 2, I-00185 Rome, Italy,\\ \& ISC-CNR, Via dei Taurini 19,
I-00185 Rome, Italy.}
\author{M. Joyce}
\affiliation{Laboratoire de Physique Nucl\'eaire et de Hautes Energies,
Universit\'e Pierre et Marie Curie-Paris 6, UMR 7585, Paris, F-75005 France.}
\author{A. Gabrielli}
\affiliation{ISC-CNR, Via dei Taurini 19,
I-00185 Rome, Italy,\\ \& SMC-INFM, Dipartimento di Fisica,
Universit\`a ``La Sapienza'', P.le A. Moro 2, I-00185 Rome, Italy.}
\author{F. Sylos Labini}
\affiliation{  "E. Fermi'' Center, Via Panisperna 89 A, Compendio del
Viminale, I-00184 Rome, Italy,\\ \& ISC-CNR, Via dei Taurini 19,
I-00185 Rome, Italy.}

\begin{abstract}    
We study, using numerical simulations, the dynamical evolution
of self-gravitating point particles in static euclidean space, 
starting from a simple class of infinite ``shuffled lattice'' initial
conditions. These are obtained by applying independently to each
particle on an infinite perfect lattice a small random displacement, 
and are characterized by a power spectrum (structure factor) 
of density fluctuations which is quadratic in the 
wave number $k$, at small $k$. For a specified form 
of the probability distribution function of the ``shuffling'' applied to 
each particle, and zero initial velocities, these initial configurations are
characterized by a single relevant parameter: the variance $\delta^2$
of the ``shuffling'' normalized in units of the lattice spacing
$\ell$. The clustering, which develops in time starting from 
scales around $\ell$,  is qualitatively very similar to that seen 
in cosmological simulations, which begin from lattices with applied 
correlated displacements and incorporate an expanding spatial background. 
From very soon after the formation of the first non-linear structures, a 
spatio-temporal scaling relation describes well the evolution of the 
two-point correlations. At larger times the dynamics of these correlations 
converges to what is termed ``self-similar'' evolution in cosmology,
in which the time dependence in the scaling relation is specified entirely by
that of the linearized fluid theory. Comparing simulations with
different $\delta$, different resolution, but identical large scale
fluctuations, we are able to identify and study features of the
dynamics of the system in the transient phase leading to this
behavior. In this phase, the discrete nature of the system explicitly
plays an essential role.
\end{abstract}    
\maketitle    
\date{today}    

\twocolumngrid  

\section{Introduction}

The problem of the evolution of self-gravitating classical particles, 
initially distributed very uniformly in infinite space, is as old as Newton. 
Modern cosmology poses essentially the same problem
as the matter in the universe is now believed to consist 
predominantly of almost purely self-gravitating particles --- so 
called dark matter --- which is, at early times, indeed very close to 
uniformly distributed in the universe, and at densities at which 
quantum effects are completely negligible. Despite the age of the problem 
and the impressive advances of modern cosmology in recent years, our 
understanding of it remains, however,  very incomplete.  
In its essentials, i.e., stripped of the full detail of current 
cosmological models, it is a simple well posed problem  
of out of equilibrium statistical mechanics\footnote{
Strictly speaking it is not actually known whether the problem
is well-controlled without a regularization of the singularity 
in the gravitational force at $r=0$
(see e.g. \cite{elskens_etal} for a recent discussion and list 
of references).
In practice, in numerical simulation, there is no intrinsic problem 
in implementing the N-body gravitational dynamics without such a 
regularisation for typical initial conditions (i.e. in which particles are not 
placed initially at the same point). In the numerical simulations 
reported here, as in  cosmological simulations, we do, however, 
use such a regularization. 
This is done solely for numerical efficiency, and the results 
analysed are tested numerically for their independence of the 
associated cut-off (see below).}. In this context, however,
it has been relatively neglected, primarily because of the
intrinsic difficulties associated with the attractive long-range nature
of gravity and its singular behavior at vanishing separation. 
In recent years there has, however, been renewed 
interest (see e.g. \cite{dauxois}) in the physics
of systems with long-range interactions, in which context 
self-gravitating systems are one of the paradigmatic 
examples (for a review, see e.g. \cite{Padmanabhan:1989gm}).
A considerable amount of work on these systems
in this context has focussed on {\it finite} systems 
(see e.g. \cite{ChavanisPRE, ispolatov_cohen, morikawa_virial,morikawa_nongauss}) ---
in which, in certain cases, some of the instruments of equilibrium 
statistical mechanics may be applied 
\footnote{We note that in \cite{devega_sanchez1,devega_sanchez2}
a treatment of infinite self-gravitating systems in the
framework of equilibrium statistical mechanics is developed
by considering a ``dilute'' infinite volume limit, in
which  $N \rightarrow \infty$ and $V \rightarrow \infty$ at 
$N/V^{1/3}={\rm constant}$, where  $N$ is the number of particles 
and $V$ is the volume (see also \cite{chavanis_2006_I} for a more 
recent discussion) . This is not the physically relevant
limit for the problem treated here, as we consider the infinite 
volume limit taken at constant density, i.e., 
$N \rightarrow \infty$ and $V \rightarrow \infty$ at $N/V ={\rm constant}$.
In this case, as discussed
further below, the system is intrinsically 
time dependent and never reaches a thermodynamic equilibrium.} --- and on more tractable 
one-dimensional models (see e.g. \cite{miller_1dgrav, miller_1dgrav_relaxation, miller_1dgrav_segregation,
miller_1dgrav_ergodicity, Tatekawa:2005nj}). 
In cosmology perturbative approaches to the problem, which treat
the very limited range of low to modest amplitude deviations from uniformity, 
have been developed (see e.g. \cite{peebles,padmanabhan}), but numerical 
simulations are essentially the only instrument beyond this regime. 
While such simulations 
constitute a very powerful and essential tool, they lack the valuable guidance 
which a  fuller analytic understanding of the problem would provide.
The dynamics of infinite self-gravitating systems is thus  
both a fascinating theoretical problem of out of equilibrium statistical 
mechanics, directly relevant both in  the context of cosmology 
and, more generally, in the physics of systems with long-range interactions.

Approaching the problem in the context of statistical mechanics, as
we do here, it is natural to start by reducing as much as possible the 
complexity of the analagous cosmological problem. We wish to
focus on the essential aspects of the problem. Thus we consider clustering 
without the expansion of the universe, and starting from particularly simple 
initial conditions. With respect to the motivation from cosmology, there
is of course a risk : in simplifying we may loose some essential elements 
which change the nature of gravitational clustering. Our results suggest 
that this is not the case. Even it were, it seems unlikely that we will
not learn something about the more complex cosmological problem in addressing
this slightly different problem.

Gravitational clustering in an infinite space --- static or expanding
-- starting from quasi-uniform initial conditions, is intrinsically a
problem out of equilibrium.
By ``quasi-uniform'' initial conditions
we mean that the initial state is a particle distribution ---
specified, we will assume, by a stochastic point process
\cite{daley} --- which has relative fluctuations at all scales,
of small amplitude above the scale characteristic of the particle
``granularity'' and decaying at infinitely large scales \footnote{It 
is also implicit in the phrase ``quasi-uniform initial conditions 
in infinite space'' that, as noted above, the infinite volume 
limit here is taken at constant particle density, rather than 
in the ``dilute'' limit studied in \cite{devega_sanchez1,devega_sanchez2}.}. 
One of the most basic 
results (see e.g. \cite{peebles,padmanabhan} and also
the appendices to this paper) about self-gravitating systems, treated in a
fluid limit, is that the amplitude of small fluctuations grows
monotonically in time, in a way which is independant of the scale.
This linearised treatment breaks
down at any given scale when the relative fluctuation at the
same scale becomes of order unity, signalling the onset of the
``non-linear'' phase of gravitational collapse of the mass in 
regions of the corresponding size. In an infinite space, in which the 
initial fluctuations are
non-zero and finite at all scales, the collapse of larger and larger
scales will continue ad infinitum. The system can therefore never
reach a time independent state, and in particular it will never reach
a thermodynamic equilibrium \footnote{This does, of course, not mean
that the instruments of equilbrium statistical mechanics are 
completely irrelevant. Saslaw (see \cite{saslaw2000} and references
therein) notably has developed a treatment of gravitational clustering 
in an expanding universe which approximates it as a ``quasi-equilbrium''
in which the thermodynamic variables evolve adiabatically with the
expansion of the Universe. Another more formal exploration of the 
usefulness of some standard equilibrium techniques can be found 
in \cite{morikawa_cluster}.}. One of the important results from
numerical simulations of such systems in the context of cosmology is,
however, that the system nevertheless reaches a kind of scaling
regime, in which the temporal evolution is equivalent to a rescaling
of the spatial variables \cite{efstathiou_88,smith}.  This
spatio-temporal scaling relation is referred to as ``self-similarity''
\footnote{Note that this term is here used in 
a different sense to that commonly ascribed to it in condensed matter
physics. In this context ``self-similarity'' usually implies that the
spatial correlations themselves have invariance properties under
rescaling (see, e.g., \cite{huang}). 
This is not necessarily the case in the present
context.}. It is observed, however, only starting from a
restricted class of simple initial conditions --- we will describe
these in further detail below --- and in the specific Einstein de
Sitter (EdS) expanding universe \cite{peebles}.  
The range of initial conditions to
which it applies has been a point of discussion in the literature, and
theoretical explanations of it typically restrict it to quite a narrow
range of such initial conditions, and strictly to the EdS expanding
universe. To see whether this kind of simple behavior is 
reproduced in the system we study, is thus a first point of
interest. It is in fact the primary focus of this paper.

One comment needs to be made about the use of a static (Euclidean)
space-time. The problem of bodies interacting by their mutual
Newtonian self-gravity in the infinite volume limit, taken at 
constant mean density, is in fact ill defined:
the force on a particle depends on how the limit is taken.  
In order to remove this ambiguity one adds a negative background to
cancel the contribution of the mean density --- the so-called ``Jeans
Swindle'' (see e.g. \cite{binney}). As discussed in \cite{gabrielli_06},
this is equivalent to taking the limit symmetrically about each particle 
on which we calculate the total gravitational force\footnote{See 
\cite{kiessling} for a very clear discussion of this issue. It is also 
shown here that addition of the negative backgound  is equivalent 
to regularizing the problem with a cosmological constant.}. Then only 
the fluctuations of the density field generate the gravitational force. 
In the context of cosmological expanding universe solutions, this ``swindle''
is unnecessary as the expansion absorbs the effect of
the mean density, and the perturbations to the comoving
particle trajectories are indeed sourced only by the fluctuations
(see, e.g., \cite{peebles}). This modification 
does not necessarily make the gravitational force well defined in general:
whether it is well defined depends on the nature of the fluctuations
in the density field at large scales. For the case of the shuffled
lattice (SL) considered here, we have studied in detail the properties
of the gravitational force in \cite{gabrielli_06}, and shown the force
to be well defined in the presence of the canceling background.

Previous works in the same spirit as this
\cite{bottaccio3,bottaccio2,Baertschiger:2004tx} have treated
primarily the very simplest initial condition one can envisage:
Poisson distributed particles with no initial velocity.  One of the
basic results which has been emphasized in these works is the role of
nearest neighbor interactions at early times in forming structures
(see also \cite{saslaw}), giving rise to non-linear density-density
correlations which
are then observed to be reproduced at larger and larger scales as time
evolves. At the same time the effects of
amplification at larger scales --- described by the fluid limit in which
the granular structure of the matter is irrelevant --- is
observed. When trying to address the basic issue of the relative
importance of these mechanisms, one runs into the limits imposed by
the simple initial conditions: in a Poisson distribution a single
parameter --- the particle density, or equivalently mean
inter-particle distance --- controls both the amplitude of 
fluctuations and the ``granularity'' of the mass distribution. This
limitation is one of the major motivations for the different class of
initial conditions we study in this work, developing further some
initial analysis of this case in \cite{Baertschiger:2004tx}: we
consider lattices subjected to small random displacements. In this
case there are now two parameters, the inter-particle distance $\ell$
and the amplitude $\Delta$ of the ``shuffling''. Given the scale free
nature of gravity it is in fact only the dimensionless combination
$\delta = \Delta/\ell$ which is physically relevant (while in the
case of Poisson initial conditions there is effectively no free
adjustable parameter).
When the dynamics of the SL is treated in the fluid limit, as we will
see, configurations with different $\delta$ may also be trivially
related. In particular we can consider systems with different $\delta$
which have different discreteness properties which are equivalent in
terms of their fluid description.  This allows us to understand
notably the aspects of the evolution of the system which can be
accounted for in a description of the dynamics in a fluid limit, and
those which require the discreteness of the system to be {\it
explicitly} taken into account. This is an important point as almost
all existing analytic results on infinite self-gravitating systems are
derived in this former limit\footnote{It is also a question which is
very relevant in the context of cosmology, as it concerns the
understanding of the discreteness effects in simulations of dark
matter, which intrinsically limit their precision. These simulations
treat the gravitational clustering of point ``macro-particles'', which
typically correspond to the order of $10^{70}$ dark matter
particles.}.  Our initial conditions are similar, but not identical,
to those used in cosmological simulations of the formation of
structure in the Universe.  In this context the initial conditions are
usually given by simple cubic lattices, perturbed by {\it correlated}
displacements, with relative displacements between nearest neighbor
particles which are small \cite{efstathiou_init}. The displacements 
are generated in reciprocal space starting from an input power 
spectrum (PS), i.e., what is usually called the ``structure factor'' 
in condensed matter physics, specifying the desired theoretical density
fluctuations.

In this paper we describe systematically basic
results on gravitational dynamics starting from
SL initial conditions. Our principal results are 
the following:

\begin{itemize}

\item Evolution from these initial conditions converges, after a
sufficient time, to a ``self-similar'' behavior, in which the
two-point correlation function obeys a simple spatio-temporal scaling
relation. The time dependence of the scaling ( i.e. the quantity
analogous to the dynamical exponent in out of equilibrium statistical
mechanics) is in good agreement with that inferred from the
linearized fluid approximation. This result is a generalization of
what has been observed, for ``redder'' initial PS
($P(k) \sim
k^n$ with $n\leq 1$), in simulations in an 
EdS universe \cite{efstathiou_88, bertschinger, smith}.

\item Between the time at which the first non-linear
correlations emerge in a given SL and the convergence
to this ``self-similar'' behavior, there is a
transient period of significant duration. During this
time, the two-point correlation function already
approximates well, at the observed non-linear scales,
a spatio-temporal scaling relation, but in which the 
temporal evolution is faster than the asymptotic evolution.
This behavior can be understood as an effect of 
discreteness, which leads to an initial ``lag'' of the 
temporal evolution at small scales.

\item Simulations with different particle numbers, but the same large
scale fluctuations (as characterized by the PS at small $k$), converge
after a sufficient time, not only to the same functional form of the
correlation function (with the self-similar behavior), but to the same
amplitude. This is further evidence that it is indeed the common large
scale fluctuations alone which determine the amplitudes of the
correlations, which are thus independent of the discreteness scale
$\ell$. At early times, however, we see manifest difference between
the systems, typically again characterized as a ``lag'' of simulations
with larger $\ell$ (and smaller $\delta$).

\item The non-linear correlations when they first develop are very
well accounted for solely in terms of two-body correlations. This is
naturally explained in terms of the central role of nearest neighbor
interaction in the build-up of these first non-linear correlations.

\item This two-body phase extends to the time of onset of the
spatio-temporal scaling, and thus the asymptotic form of the
correlation function is already established to a good approximation at
this time. We briefly discuss the significance of this quite
surprising finding.

\end{itemize}

The paper is organized as follows. In the next section we briefly
define a SL distribution and introduce the main statistical quantities
we use in the analysis and their estimators. We discuss the numerical
simulations and their analyses in Sec.\ref{sec3}. Finally in
Sec.\ref{sec4} we summarize our mains results and conclusions, and 
briefly discuss some of the many open problems which remains for future 
investigation.


\section{Shuffled Lattices and Statistical Quantities}
\label{sec2} 

We firstly describe (Sec.\ref{sec:charact}) the class of initial
conditions we study. In Sec.\ref{sec:stat} we define the statistical
quantities we will use to characterize the correlations, and in
Sec.\ref{sec:est} we specify how we estimate these quantities in our
simulations.

\subsection{Definition of a Shuffled Lattice}
\label{sec:charact}

We use the term SL to refer to the infinite point distribution
obtained by randomly perturbing a perfect cubic lattice: each particle
on the lattice, of lattice spacing $\ell$, is moved randomly
(``shuffled'') about its lattice site, each particle {\it
independently} of all the others.  A particle initially at the lattice
site $\ve R$ is thus at $\ve x(\ve R) = \ve R + \ve u(\ve R)$, where
the random vectors $\ve u(\ve R)$ are specified by the factorised joint
probability density function
\be
{\cal P}[\{ \ve u(\ve R) \}]= \prod_{ \ve R } p(\ve u(\ve R)). 
\ee
The distribution is thus entirely specified by $p(\ve u)$, the 
probability density function (PDF) for the displacement
of a single particle.

\begin{figure}
\includegraphics[width=0.35\textwidth]{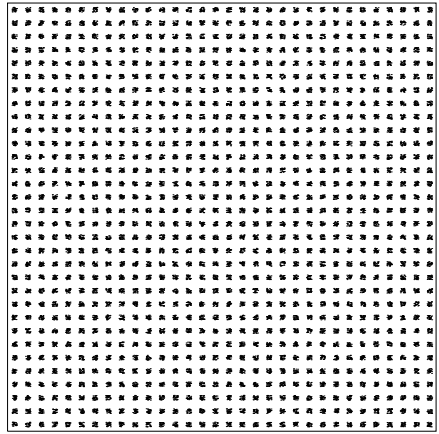}
\caption{Projection on the $z=0$ plane of a SL with $32^3$ particles
and $\delta=0.177$. Due to the random shuffling with the given PDF
each lattice ``chain'' parallel to the $z$-axis projects onto a small 
square. \label{fig:shuffled_lattice}}
\end{figure}

In this paper we will study evolution from SL with the 
following specific PDF \footnote{We will 
discuss in the conclusions section the importanc of this 
specific choice for the PDF.}:
\begin{equation}
p(\ve u) = 
\begin{cases}
  (2\Delta)^{-3} & \text{if } \ve u \in [-\Delta,\Delta]^3 \ ,
  \\ 
  0 & \text{otherwise.}
\end{cases}
\label{eq:pu}
\end{equation}
Each particle is therefore moved randomly in a cube of side $2\Delta$
centered on the corresponding lattice site
(Fig.~\ref{fig:shuffled_lattice}). 
Taking $\Delta\to 0\,$, at fixed $\ell$, one thus obtains a
perfect lattice, while taking $\Delta\to \infty$ at fixed $\ell$, one
obtains an uncorrelated Poisson particle configuration \cite{book}.
Given Eq.~(\ref{eq:pu}), the \emph{shuffling parameter}
$\Delta^2$ also gives the variance of the shuffling,
i.e.,
\begin{equation}
\Delta^2 = \int d^3 u   {\ve u}^2 p (\ve u ).
\end{equation}
Our SL configurations are therefore specified by two parameters: the
lattice constant $\ell$ and the shuffling parameter $\Delta$.  An
alternative convenient characterization is given by $\ell$ and the
adimensional ratio $\delta \equiv \Delta/\ell$.  We will refer to the
latter as the \emph{normalized shuffling parameter}. It is thus the
square root of the variance of the shuffling in units of the lattice
spacing.  

In what follows we consider not an infinite SL, but a finite SL of $N$
particles in a cubic box of size $L=N^{1/3} \ell$ (see
Fig.~\ref{fig:shuffled_lattice}). We will consider the specific case
of a {\it simple cubic lattice}, in which thus the mean number density
of particles is thus $N/L^3=n_0=\ell^{-3}$. Further we will assign to
all particles the same mass $m$, so that the average mass density is
simply $\rho_0 = m n_0$.

In Table~\ref{tab:sl_summary} we list the various relevant parameters
of the SL considered as initial conditions of the N body simulations
(NBS) which we report here. 
We will explain below the criteria used for these choices.
\begin{table}
\begin{ruledtabular}
\begin{tabular}{cccccccc}
Name & $N^{1/3}$ & $L$ & $\ell$ & $\Delta$ & $\delta$ & $m/m_{64}$ \\ 
\hline 
SL64 & 64 & 1& 0.015625& 0.015625 & 1       & 1     \\ 
SL32 & 32 & 1&0.03125  & 0.0553   & 0.177   & 8     \\ 
SL24 & 24 & 1&0.041667 & 0.00359  & 0.0861  & 18.96 \\ 
SL16 & 16 & 1&0.0625   & 0.00195  & 0.03125 & 64    \\ 
\hline 
SL128& 128&2 &0.015625 & 0.015625 & 1       & 1     \\ 
\end{tabular}
\end{ruledtabular}
\caption{Details of the SL used as initial conditions in the
  simulations reported in this paper. $N$
  is the number of particles, $L$ is the box size, $\ell$ the lattice
  constant and $\Delta$ ($\delta$) the (normalized) shuffling parameter.
  The mass $m$ of the particles is expressed in unit of that in SL64,
  i.e., $m_{64}$. In the units chosen, the mass density in all these
  systems is $\rho_0 = Nm/L^3=1$.  Note that SL64 and SL128 are ``more
  shuffled'' than all the others (i.e. larger shuffling parameter)
  while SL16 is the one which is the closest to a perfect cubic
  lattice. Note that SL128 differs only from  SL64 by the size of
the box.
  \label{tab:sl_summary} }
\end{table}

\subsection{Statistical Characterization of Correlation Properties} 
\label{sec:stat}

The microscopic number density function for any particle distribution
is given by
\begin{equation}
n(\ve x) = \sum_{i=1}^N \delta_\text{D}\bigl(\ve x- \ve x_i\bigr) \;,
\end{equation}
where $\ve x_i$ is the position of the $i$-th particle, 
$\delta_\text{D}$ is the Dirac delta function and the sum 
is over the $N$ particles of the system.

\subsubsection{The two-point correlation function}
For a system such as we consider here, in which the mean density
is well defined and non-zero, it is convenient to define the
density contrast: 
\begin{equation}
\delta(\ve x) = \frac{n(\ve x) - n_0}{n_0} \;.
\end{equation}
In order to characterize the two-point correlation properties of the
density fluctuations, one can then use the reduced two-point 
correlation function:
\begin{equation} 
\tilde \xi(\ve r)= \langle \delta(\ve x+\ve r)\delta(\ve x)\rangle
 \;,
\end{equation}
where $\langle ... \rangle$ is an ensemble average, i.e., an average
over all possible realizations of the system. In a
distribution of discrete particles $\tilde{\xi}(\ve r)$ 
always has a Dirac delta function singularity at 
$\ve r=0$, which it is convenient to separate
by defining $\xi(\ve r)$ for $\ve r \neq \ve 0$
(the ``off-diagonal'' part) ~\cite{book}:
\begin{equation}
\tilde{\xi}(\ve r) = \frac{1}{n_0} \delta_D(\ve r) + \xi(\ve r) \ . 
\label{eq:xigeneral}
\end{equation}
It is useful also to note that one can write
\begin{equation}
\xi(r) = \frac{\langle n(r)\rangle_p }{n_0}-1 
\;.
\label{eq:xi_rdf}
\end{equation}  
where $\langle n(r)\rangle_p$, the 
\emph{conditional average density}, is the 
(ensemble) average density of points in an 
infinitesimal shell at distance $r$ {\it from an 
occupied point} \footnote{For the more general case
of non-uniform distributions, such as fractal particle distributions
\cite{book}, in which $n_0$ is 
zero, this is the basic statistical quantity for 
the characterization of two-point correlation
properties, rather than $\tilde{\xi}(r)$ (which is
then not defined).}. We will make use of this 
relation is estimating $\xi(r)$ below.

In the evolved self-gravitating systems we study
$\xi(r)$ will invariably be a monotonically decreasing
function of $r$. It is then natural to define
a scale $\lambda$ by  
\begin{equation}
\xi(\lambda) = 1  
\label{eq:homoscale}
\end{equation}
which separate the regime of weak correlations
(i.e. $\xi(r) \ll  1$) from the regime of strong 
correlations (i.e. $\xi(r) \gg  1$).
In the context of gravity these are what are referred
to as the {\it linear} and {\it non-linear} regimes,
as a linearized treatment of the evolution of
density fluctuations is expected to be valid in the 
former case.
Given the form of Eq.~(\ref{eq:xi_rdf}) it is clear
that $\lambda$ as defined by Eq.~(\ref{eq:homoscale})
is an appropriate definition of the
\emph{homogeneity scale} of the system. This scale 
gives then the typical size of strongly clustered
regions.

The exact analytic expression for $\xi(r)$ in a SL is given in
\cite{gabrielli_06}.  We do not reproduce it here as it is a
complicated expression, which we will not in fact make use of.  In
our case, as in the case of a perfect lattice and a Poisson
distribution (which, as we have noted correspond to specific limits of
the SL) $\xi(r) <1$ everywhere: there is no strong clustering. In such
a situation the homogeneity scale is of order of the average distance
between nearest neighbors (NN), which we will denote by
$\Lambda$. Thus when we refer to the homogeneity scale we will mean
$\Lambda$ in absence of non-linear clustering and the scale given by
Eq.(\ref{eq:homoscale}) otherwise.

\subsubsection{The mass variance}

For particle distributions with a well defined average density
it useful also to consider an integrated quantity such as the 
normalized variance of particle number (or mass)  in 
spheres, defined as
:
\begin{equation}
\label{eq:variance} 
\sigma^2(r) = \frac{\langle N^2(r)\rangle - \langle N(r)\rangle^2
}{\langle N(r) \rangle^2} 
\end{equation}
where $N(r)$ is the number of particles in a sphere of radius $r$.
Then $\sigma^2(r)$ can be used, in a manner similar to that described
above for $\xi(r)$, to distinguish a regime of  large {\it fluctuations} 
from a regime of small fluctuations. It is simple to find the explicit
expression for $\sigma^2(r)$, which gives it as a double integral
of $\xi(r)$ \cite{book}. 

One can show that the normalized variance in real-space spheres
defined in Eq.~(\ref{eq:variance}) behaves in a SL as $\sigma^2(r)
\propto r^{-4}$ at large $r$, compared to $\sigma^2(r) \propto r^{-3}$
in a Poisson distribution \cite{glasslike,book}. The behavior
$\sigma^2(r) \propto r^{-4}$ is in fact the fastest possible
decay of this quantity \cite{book}. This means that
the SL belongs to the class of distributions which may be termed 
\emph{super-homogeneous} \cite{glasslike} (or \emph{hyper-uniform}
\cite{torquato}). Such systems have mass fluctuations which are 
depressed with respect to those in an uncorrelated Poisson
distribution.

\subsubsection{The Power Spectrum}           

Since we consider distributions which are periodic in
a cube of side $L$, we can write the density contrast as a 
Fourier series:
\begin{equation}
\delta(\ve x) = \frac{1}{L^3}\sum_{\ve k} \exp(i\ve k\cdot \ve x)
\,\tilde\delta(\ve k)
\label{eq:fourierdelta}
\end{equation}
with $\ve k \in \left\{ (2\pi/L) \ve n \, |\, \ve n \in
\mathbb{Z}^3\right\}$.  The coefficients 
$\tilde\delta(\ve k)$ are given by
\begin{equation}
\tilde\delta(\ve k) =\int_{L^3} \delta(\ve x)
\exp(-i\ve k\cdot \ve x) \ \D[3] \ve x \ . 
\end{equation}
The PS of a particle distribution\footnote{We use here the term
for this quantity commonly employed in cosmology, rather 
than ``structure factor'' which is more habitual in the 
context of condensed matter and statistical physics. 
Note also the normalisation, which corresponds to
$P( \ve k \rightarrow \infty ) \rightarrow \frac{1}{n_0}$,
rather than unity.}  
is then defined as (see e.g. \cite{andrea,book})
\begin{equation}
P(\ve k) = \frac{1}{L^3}\langle | \tilde\delta(\ve k) |^2 \rangle \;.
\label{eq:pktheo}
\end{equation}
In distributions which are statistically homogeneous, which is the case
here\footnote{For the lattice and SL the ensemble average
is defined over the set of lattices rigidly translated  by an arbitrary vector
in the unit cell.}, the PS  and  reduced two-point correlation
function $\tilde{\xi} (\ve r)$ are a Fourier conjugate pair. 

The exact expression for the PS of a SL is simple to derive.
One finds (see \cite{andrea,book}) 
\begin{equation}
 P(\ve k) = \frac{1-|\tilde{p}(\ve k)|^2}{n_0} + L^3 \sum_{\ve n}
 |\tilde{p}(\ve k)|^2 \,  \delta_\text{K}( \ve k , \ve n \frac{2 \pi}{\ell}) 
\label{eq:exactPS}
\end{equation}
where $\tilde p(\ve k)$ is the Fourier transform of the 
PDF for the displacements $p(\ve u)$ (i.e. its characteristic 
function),  and $\delta_\text{K}$ is the three-dimensional 
Kronecker symbol. For the specific  $p(\ve u)$ given
in Eq.~\eqref{eq:pu} we have 
\begin{equation}
|\tilde p(\ve k)|^2=
\prod_{i=x,y,z} \frac{\sin^2(k_i \Delta)}{( k_i \Delta)^2} \;.
\label{eq:charaSL} 
\end{equation}
Inserting this expression in Eq.~(\ref{eq:exactPS}) one obtains an
exact explicit analytic expression for the PS of a SL in terms of the
two parameters $\ell$ and $\Delta$. It is simple to verify that taking
$\Delta/\ell =\delta \rightarrow \infty$, at fixed $\ell$, one obtains
$P(k)=1/n_0$ (as expected, since one obtains in this limit a Poisson
distribution). Further one always approaches this same behavior (as
required \cite{book}) in the limit $k \to \infty$. Further, at small
$k$ (i.e.  $ k \ll 2 \pi/\ell$), we obtain
\begin{equation}
P(\ve k) \approx \frac{|\ve k|^2 \Delta^2}{3n_0}\,.   
\label{eq:pk}
\end{equation}
We note that this result can actually be found 
(see \cite{andrea,book}) directly from Eq.~(\ref{eq:exactPS}),
without assuming a specific form for $p(\ve u)$.
One need only assume that $p(\ve u)$ has a finite
variance, equal to $\Delta^2$. {\it Thus the small $k$
behavior of the PS of the SL does not depend on
the details of the chosen PDF for the displacements,
but only on its (finite) variance}.
\footnote{Note that the 
behavior $\lim_{\ve k \to \ve 0} P( \ve k)=0$ is an equivalent way
of stating the property of super-homogeneity of the distribution
\cite{book}.}.

Finally note that the mass variance can actually be expressed simply
as an integral in reciprocal space of the PS multiplied by an
appropriately normalized Fourier transform $\tilde W(k;r)$ of the
spherical window function, being 0 outside the sphere and 1 inside it
\cite{book}:
\be
\sigma^2(r)= \frac{1}{(2\pi)^3 }\int d^3 k P(k)|\tilde W(k;r)|^2 \;,
\label{sigma-ps}
\ee

\subsubsection{The nearest neighbor distribution}

A very useful and simple statistical quantity which characterizes 
small-scale clustering properties of a particle distribution is the 
nearest neighbor (NN)
PDF $\omega(r)$. It is the probability density for the distance
between a particle and its NN \cite{book}, i.e., $\omega(r)dr$ gives
the probability that a particle has its NN at a distance
in $[r, r+dr]$. If one neglects correlations of
any order higher than two, it is simple to show that
$\omega(r)$ is related to the conditional density 
$\langle n(r)\rangle_p$ (and thus to $\xi(r)$, given
Eq.~\eqref{eq:xi_rdf}) through
\footnote{The relation
follows if one assumes that the probability of finding a 
particle in $[r, r+dr]$ given that there is a point at $r=0$
is the same whether the condition that there be no 
other point in  $[0, r]$ is imposed or not.}
\begin{equation} 
\label{eq:omega1} 
\omega(r) \, \D r = \left( 1 - \int_0^r\omega(s)\, \D s \right) \cdot
 4\pi r^2  \langle n(r)\rangle_p  \D r  \ , 
\end{equation}
This relation will be very useful to us here because it is valid
in particular when clustering is dominated, at early times, 
by individual pairs of particles falling toward each other.


\subsection{Estimation of Statistical Quantities} 
\label{sec:est}

In order to estimate $P(\ve k)$ and $\xi(\ve r)$ in a given particle
configuration, i.e., in a single realization of the evolved
SL, we calculate averages in spherical shells in real or 
reciprocal space. This means that we consider only
the dependence of these quantities on the modulus of their arguments
and we will therefore use the notation $P(k)$ and $\xi(r)$ in the rest
of the paper. 

The PS is obtained from $\tilde{\delta}(\ve k)$ by means of the
relation\footnote{For simplicity in this paper we use the same symbol
for the ensemble average quantity and for its estimator.}
\begin{equation}
P(k) \approx \frac{1}{N(k)} \sum_{ k\leq  |\ve k'| 
\leq
k + \delta k}  | \tilde{\delta}(\ve k')|^2
\label{eq:ps_estimated}
\end{equation}
where $N(k)$ is simply the number of vectors $\ve k'$ considered in the
sum. Note that to speed up the calculations, not all the vectors $\ve
k'$ for a given modulus are taken into account: at large $\ve k$ the
density of vectors considered is smaller than at small $\ve k$. 

The function $\xi(r)$ is estimated by first calculating $\langle
n(r)\rangle_p$ (see Eq.~\ref{eq:xi_rdf}) ~\cite{book}. As already
mentioned the latter gives the average density in a spherical shell of
radius $r$, and thickness $\delta r\ll r$, centered on an occupied point.
Thus we estimate it as
\begin{equation}
\langle n(r)\rangle_p 
\simeq 
\frac{1}{V(r,\delta r) N_c} \sum_{i=1}^{N_c}
N_i(r)
\label{eq:estim_gr}
\end{equation}
where $N_i(r)$ is the number of particles\,\footnote{We use periodic
boundary conditions in this estimation (as in the simulations).}  
in the spherical shell of radii
$r,\ r+\delta r$, volume $V(r,\delta r)$, centered on the $i^{th}$
particle of a subset of $N_c \le N$ particles randomly chosen among
the $N$ particles of the system.

The mass variance can be simply estimated by 
\begin{equation}
\sigma^2(r) 
\simeq 
\frac{1}{ \langle N(r) \rangle^2}   \frac{1}{N_c-1} \sum_{i=1}^{N_c} 
\left( N_i(r) - \langle N(r) \rangle\right)^2
\label{eq:estim_gr2}
\end{equation}
where $N_i(r)$ is the number of particles contained in the $i^{th}$
(with $i=1..N_c$) {\it randomly placed} sphere of radius $r$ and
$\langle N(r) \rangle$ its average.

Finally the NN distribution $\omega(r)$ is computed directly
by pair counting.


\section{Gravitational Clustering  in a Shuffled Lattice:
Results from Numerical Simulations}
\label{sec3} 

\subsection{Details of Numerical Simulations}

We have performed a set of numerical simulations using the freely
available code \textsc{Gadget} \cite{gadget,gadget_paper}. This code,
which is based on a tree algorithm for the calculation of the force,
allows one to perform simulations in an infinite space, using the
Ewald summation method~\cite{hernquist}.  The potential used is {\it
exactly} equal to the Newtonian potential for separations greater than
the softening length $\varepsilon$, and regularized at smaller scales.
For what concerns the integration parameter we have performed several
tests to check the stability of the results at the level of numerical
precision we consider in this work \footnote{In order to test the
numerical accuracy of the simulations we have also compared the early times
evolution with the prediction  of the linearised treatment
of the early time evolution, as described in detail in \cite{marcos_06}.}.

We have considered as initial conditions the set of five SL described
in Table~\ref{tab:sl_summary}.  We now explain the reasons for our
choices of the parameters given.

Firstly it is important to note that, in the limit of the pure
(i.e. un-softened) gravitational evolution of an infinite SL, there is
only {\it one} parameter which can change the dynamical evolution
non-trivially
\footnote{By ``dynamical evolution'' we mean the ensemble average
properties of the clustering etc.. We thus suppose that this average
is recovered in a single realization of an infinite volume system,
i.e., that spatial ergodicity applies. The unaveraged dynamical
evolution will of course vary in detail from one realization to
another.}.  This is $\delta$, the normalized shuffling parameter
(i.e. normalized in units of the lattice spacing $\ell$). Because
gravity has no preferred length scale the gravitational dynamics of
two infinite SL with the same $\delta$, but different lattice spacing
$\ell$, can be trivially related: a rescaling of length scales is
equivalent to a rescaling of time, so that the configurations of one
can be mapped at any time onto the configuration of the other at a
different time. The same is true for changes of the mass of the
particles: two SL with the same $\delta$, the same $\ell$, but
different particle masses, are related by a simple scaling of the time
variable. Indeed any two SL with the same $\delta$, are strictly
equivalent to one another in time if they are related to one another
by any simultaneous scaling of $\ell$ {\it and} the particle mass
which leaves their mass density $\rho_0$ fixed.

For {\it softened} gravity in a {\it finite} box, the same length
scale transformations can relate trivially different SL with the same
$\delta$. In this case the relevant parameters to distinguish two SL
evolved in these simulations are thus {\it three}, which we can take
as $\delta$, $\ell/L=N^{1/3}$ and $\epsilon/L$.

We have chosen our (arbitrary) units of length, mass and time as
follows. Our unit of length is given by the box side of the SL64
simulation and our unit of mass by the particle mass in this same
simulation. A natural choice for the unit of time is the so called
\emph{dynamical time}, defined as
\begin{equation}
\tdyn \equiv \frac{1}{\sqrt{4\pi G \rho_0}}.
\end{equation}
As unit of time here we have made a slightly different
choice of the pre-factor, with $\tdyn=1.092$
\footnote{This corresponds to time in units of $1000$ seconds 
for a mass density of $1$g.cm$^{-3}$.}.

In the ``reference'' SL64 simulation we have chosen $\delta=1$.  Our
choices of the parameters for the other five simulations can be
understood as follows:

\begin{itemize}
\item  The particle masses are chosen so that the mass density is constant.
Thus the dynamical time is the same in all simulations, which is
convenient for comparison, as we will see, as this is the unique 
time scale of these systems in the fluid limit.

\item The softening $\varepsilon$ is the same in all the simulations.
We have chosen $\varepsilon = 0.00175$ in our units, which means
that  it is, in all the simulations, significantly smaller (at least a factor 
of ten) than $\Lo$, the initial average distance between NN 
\footnote{The smallest value of $\Lo$ is that in SL64 where it
is equal to $0.55/64\approx 0.0086$ as in a Poisson distribution 
with  the same number density \cite{book}.}.

\item The box size is the same in all but one simulation.
This latter simulation (SL128), which is the biggest one,
is used to test the accuracy with which our results are
representative of the infinite volume limit (at fixed
mass and particle density).  Thus it is chosen to have
the same parameters to SL64, differing only in its volume 
(which increases by a factor of 8).

\item  For each of the four other SL simulations we change 
the number of particles $N$, which fixes $\ell$. We have then chosen
$\delta$ {\it so that the PS at small $k$ has the same
amplitude}. From Eq.~\eqref{eq:pk} it it easy to see that this
requires, in our chosen units,
\begin{equation}
\frac{\Delta^2}{n_0} \equiv  \delta^2 \ell^5 
=\left( \delta^2 \ell^5 \right)_{\text{SL64}} = 64^{-5}  \ ,
\end{equation}

\end{itemize}

\begin{figure}
\includegraphics[width=0.5\textwidth]{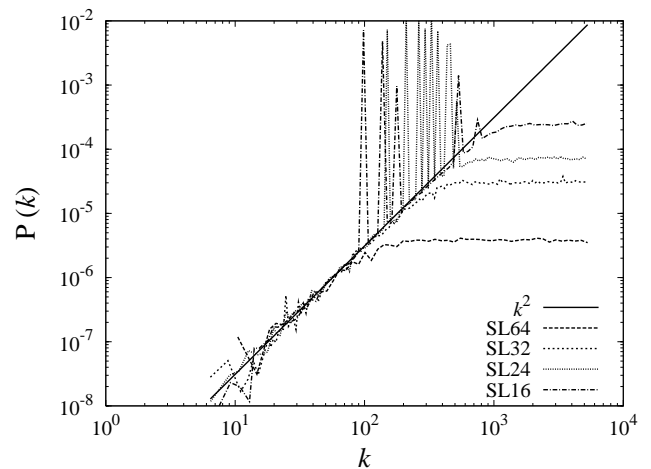}
\caption{The PS (averaged in spherical shells) of the SL configurations
specified above in Tab.~\ref{tab:sl_summary} as a function of the 
modulus of $\ve k$. The
solid line is the theoretical ($\propto k^2$) behavior for small $k$
given by Eq.~\eqref{eq:pk}. At large $k$, the four PS are equal to
$1/n_0$, with the corresponding value of $n_0$. The peaks arise from
the second term in Eq.~\eqref{eq:exactPS}. The four arrows show the
different Nyquist wave-numbers multiplied by two for the SL configurations
in order of increasing number density from left to right: this
corresponds to the expected location of the first peak in each case. 
Note that, as we have discussed in
Sect.~\ref{sec:est}, not all the vectors $\ve k$ are considered in the
estimation of the PS and therefore only a subset of all the peaks is
detected (each peak corresponds to a very narrow band of $\ve k$ so it can be
easily missed).
\label{fig:pk_t0}}
\end{figure}

The PS of the SL described in Table~\ref{tab:sl_summary} are shown in
Fig.~\ref{fig:pk_t0}. We see, up to statistical fluctuations, that the
spectra are indeed of the same amplitude at small $k$.  Note that the
Nyquist frequency $k_N=\pi/\ell$ in $k$-space translates to the right
with increasing particle number.

The particles are assigned {\it zero velocity} in the initial
conditions (at $t=0$), and, as has been underlined, the simulations
are performed in a static Euclidean universe, i.e., without expansion
or non-trivial spatial curvature.  We have run the simulations SL16, SL24,
SL32 and SL64 up to about time 6 and the SL128 up to time 8 as for
longer times the simulations begin to be dominated by a single
non-linear structure, a regime in which we are not interested since it
is evidently strongly affected by finite size effects.


\subsection{Results}

In this section we analyze the results of the numerical evolution of
the SL described in Table~\ref{tab:sl_summary} in terms of the
statistical quantities discussed above. In the first two subsections
we restrict ourselves to the study of the evolution of SL128, i.e.,
the largest simulation we have run. This is then our reference point
with which we compare the evolution of the other initial conditions.

\subsubsection{Evolution of the Power Spectrum} 
\label{ps-evolution}

\begin{figure}
\includegraphics[width=0.5\textwidth]{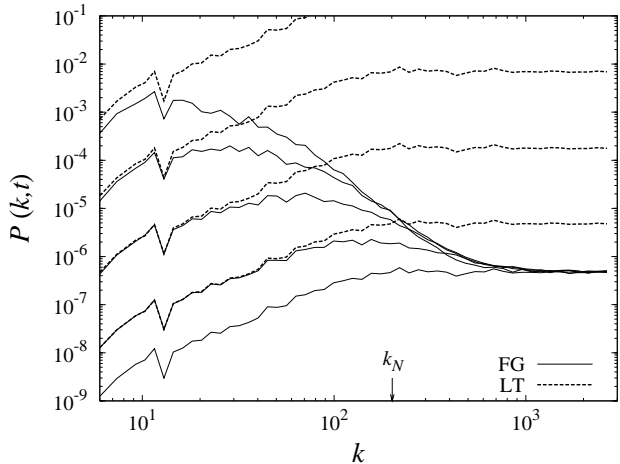}
\caption{Evolution of the PS in SL128 (solid lines --- label FG): the
  curves are for time equal to 0,2,4,6,8 (from bottom to up).  The
  dashed lines labeled with LT show the predictions of fluid linear
  theory, i.e., Eq.~\eqref{eq:pk_evolution_linear} with $P( k,0)$
  measured in the simulation at $t=0$ for the same time steps. The
  arrow labeled ``$k_N$'' shows the value of the corresponding Nyquist
  frequency $k_N = \pi/\ell$.
\label{fig:pk_SL128} }
\end{figure}

The evolution of the PS in SL128 estimated by using
Eq.~\eqref{eq:ps_estimated} is shown in Fig.~\ref{fig:pk_SL128}. Along
with the numerical results is shown the prediction for the evolution
of the PS given by the linearized fluid theory (see App.~\ref{app-a}):
\begin{equation}
P(\ve k,t) = P(\ve k,0) \cosh^2\left(t/\tdyn \right) \ .
\label{eq:pk_evolution_linear}
\end{equation}

We observe that:

\begin{itemize}

\item 
The linear theory prediction describes the evolution very accurately
in a range $k < k^{*}(t)$, where $k^{*}(t)$ is a wave-number which
decreases as a function of time. This is precisely the qualitative
behavior expected as linear theory is expected to hold only above a
scale which, in real space, increases with time, and, in reciprocal
space, decreases with time. We note that at $t=6$ only the very
smallest $k$-modes in the box are still in this {\it linear
regime}, while at $t=8$ this is no longer true.
We will discuss below a more precise quantification of the
validity of the linearized approximation.

\item
At very large wave-numbers ($k > k_N$) the PS remains equal to its
initial value $1/n_0$. This is simply a reflection of the necessary
presence of shot noise fluctuations at small scales due to the
particle nature of the distribution. We note that the value of $k$
at which this behavior is attained increases somewhat from its
initial value and then remains roughly stable. We will comment further
on the significance of this fact below.

\item 
In the intermediate range of $k$, i.e., $k^*(t) < k \ltapprox k_N$, the
evolution is quite different, and {\it slower}, than that given by
linear theory. This is the regime of {\it non-linear clustering}.
\end{itemize} 

These results concerning the validity of linear theory at sufficiently
small $k$, and in a range which decreases as a function of time, are
completely in line with what is observed in cosmological simulations,
in an expanding universe (see e.g. \cite{miller_83}). In this context
simulations typically start from lattices with correlated perturbations
representing spectra which are much ``redder'' than $P(k) \sim k^2$,
typically $P(k) \sim k^n$ with $-3 < n <0$. That the same behavior is
seen in a static universe for this ``bluer'' PS is, however,
expected. Indeed, on the basis of simple considerations (see e.g.
\cite{peebles}) --- which do not make use of the expansion of the
universe --- about the long-wavelength (i.e. small $k$) perturbations
generated by non-linear motions on small scales, one anticipates that
linear theory should be valid at small $k$ for any initial PS with
$P(k) \sim k^n$ and $n <4$. The reason is that such non-linear motions,
which preserve locally mass and center of mass, can generate at most a
PS at small $k$ with the behavior $P(k) \sim k^4$.

\subsubsection{Evolution of the Two-Point Correlation Function}

\begin{figure}
\psfrag{ell}{$\ell$}
\includegraphics[width=0.5\textwidth]{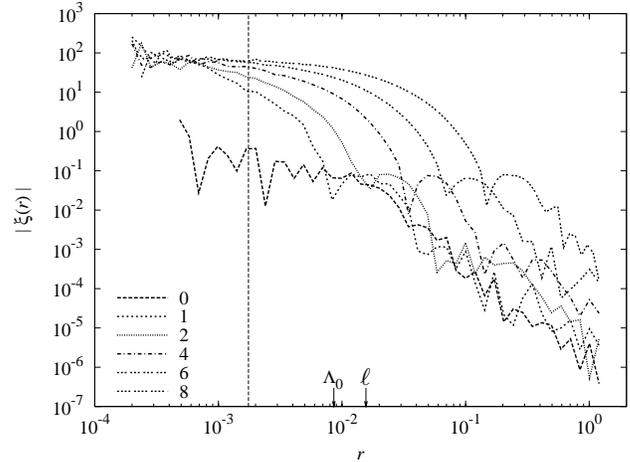}
\caption{Behavior of the absolute
value of the correlation function $|\xi(r)|$ in SL128 at times
$t=0,1,2,4,6,8$. Note that values such that $|\xi(r)| < 0.01-0.1$ 
are below the level 
of noise of the estimator  estimated by using the normalized
variance in spherical shells (see text): This gives a limit below
which the noise in the estimator dominates over the signal. The arrows
shows the value of the lattice spacing $\ell$ and the initial average
distance between nearest particles $\Lambda_0$, while the dotted
vertical line corresponds to the smoothing $\varepsilon$.
}
\label{fig:xi}
\end{figure}

We consider now the evolution of clustering in real space, as
characterized by the reduced correlation function $\xi(r)$. We focus
again on SL128. In Fig.~\ref{fig:xi} is shown the evolution of the
absolute value $|\xi(r)|$ in a log-log plot. In the figure is shown
also, for comparison, at large scales, the level of the typical
fluctuations expected in the estimator of $\xi(r)$ \footnote{This
estimate is obtained by assuming that the variance in the shells
employed in the estimator decay as
$\sqrt{\sigma_{\text{shell}}^2(r)/N_c}\propto r^{-2}$, where
$\sigma_\text{shell}^2(r)$ is the variance in shells, defined
analogously to that in spheres [cf. Eq.~\eqref{eq:variance}], and
$N_c$ is the number of centers used to calculate $\xi(r)$ [cf.
Eq.~\eqref{eq:estim_gr}].}. This indicates that, at larger
separations, the noise in the estimator is expected to dominate 
over any underlying physical correlation which may be present.

\begin{figure}
\includegraphics[width=0.5\textwidth]{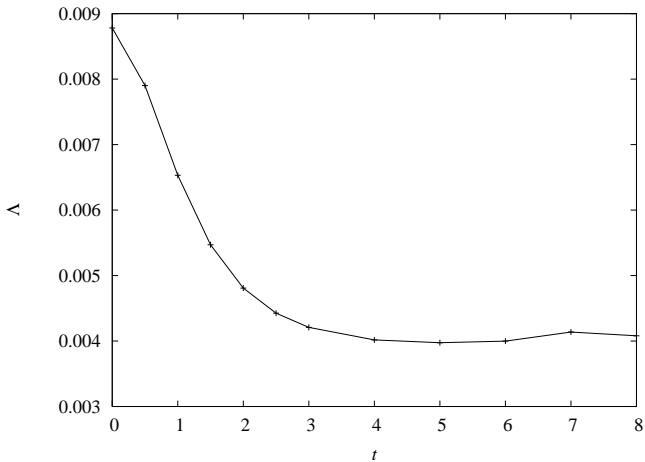}
\caption{Evolution in time of $\Lambda(t)$, the average distance between 
nearest  neighbors, in SL128. It decreases at early times and
then stabilizes at $\Lambda \approx 2 \varepsilon$. }
\label{fig:sl128lambda}
\end{figure}

We observe that:

\begin{itemize}
\item Starting from $\xi(r)\lesssim 1$ everywhere, non-linear correlations
(i.e. $\xi(r)\gg 1$ ) develop first at scales smaller than the
initial inter-particle distance $\Lo$. 

\item After two dynamical times the clustering
develops little at scales below $\varepsilon$. The clustering at these
scales is characterized by an approximate ``plateau'' at $\xi(r) \sim
10^2$. This stabilization of the system at small scales is also
evident in Fig.~\ref{fig:sl128lambda}, which shows the evolution of
the mean distance between nearest neighbor particles as a function of
time. The stabilization in time of the scale in $k$ space at which the
PS reaches its asymptotic (constant) value, which we observed above,
is just the manifestation in reciprocal space of this same
behavior. 

\item At scales larger than $\varepsilon$ the correlations
grow continuously in time at all scales, with the scale of
non-linearity [which can be defined, as discussed above, by
$\xi(\lambda)=1$] moving to larger scales.

\end{itemize}

From Fig.\ref{fig:xi} it appears that, once significant non-linear
correlations are formed, the evolution of the correlation function
$\xi(r)$ can be described, approximately, by a simple ``translation''
in time. This suggests that $\xi(r,t)$ may satisfy in this regime a
spatio-temporal scaling relation:
\begin{equation}
\xi(r,t) \approx \Xi\left( r/R_s(t) \right)\,, 
\label{eq:rescaling}
\end{equation}
where $R_s(t)$ is a time dependent length scale which we discuss in
what follows.
\begin{figure}
\includegraphics[width=0.5\textwidth]{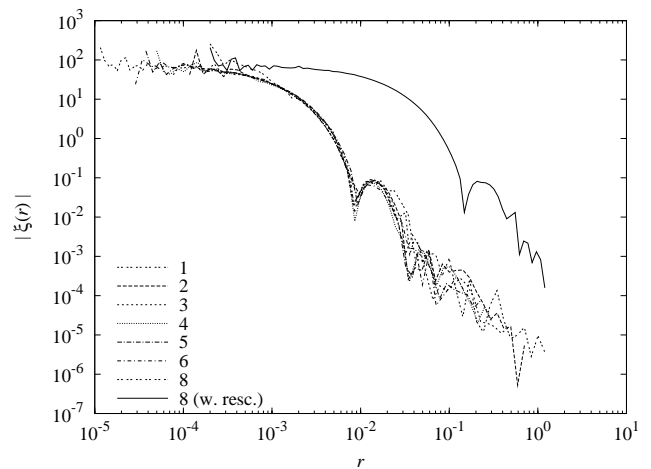}
\caption{Collapse plot of $\xi(r,t)$: for each time $t>1$ we 
have rescaled the $x$-axis by a time-dependent factor to collapse all
the curves (dashed ones) to that at time $t=1$. We have added for
comparison $\xi(r,t=8)$ without rescaling (``w. resc.'', continuous
line).}
\label{fig:collapse}
\end{figure}
In order to see how well such an ansatz describes the evolution, 
we show in
Fig.~\ref{fig:collapse} an appropriate ``collapse plot'': $\xi(r,t)$
at different times is represented with a rescaling of the $x$-axis by
a (time-dependent) factor chosen to superimpose it as closely as
possible over itself at $t=1$, which is the time from which the
``translation'' appears to first become a good approximation.  We can
conclude clearly from Fig.~\ref{fig:collapse} that the relation
\eqref{eq:rescaling} indeed describes very well the evolution, down to
separations of order $\varepsilon$, and up to scales at which
the noise dominates the estimator.

\begin{figure}
\includegraphics[width=0.5\textwidth]{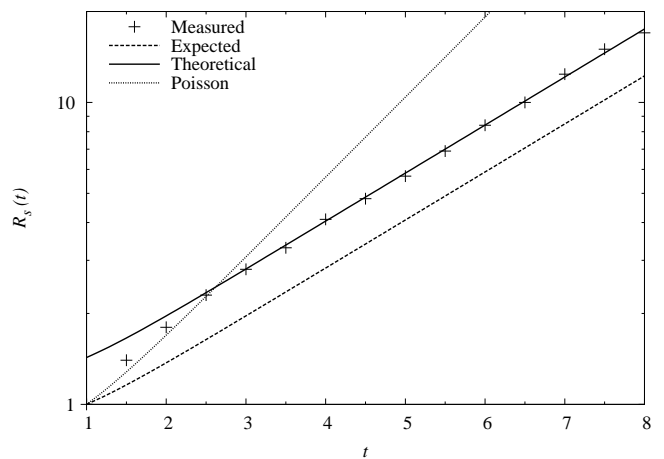}
\caption{Evolution of the function $R_s(t)$ in SL128
(points) compared with its prediction (``expected''
and ``theoretical'') for different values of the 
time scale of onset of self-similarity
(see text for details). Both lines corresponds to 
$R_s(t)\propto \exp[ (2/5) t/\tdyn]$. Also shown is 
the corresponding prediction for Poisson initial conditions,
$R_s(t)\propto \exp[ (2/3) t/\tdyn]$.}
\label{fig:Rst}
\end{figure}

\begin{figure}
\includegraphics[width=0.5\textwidth]{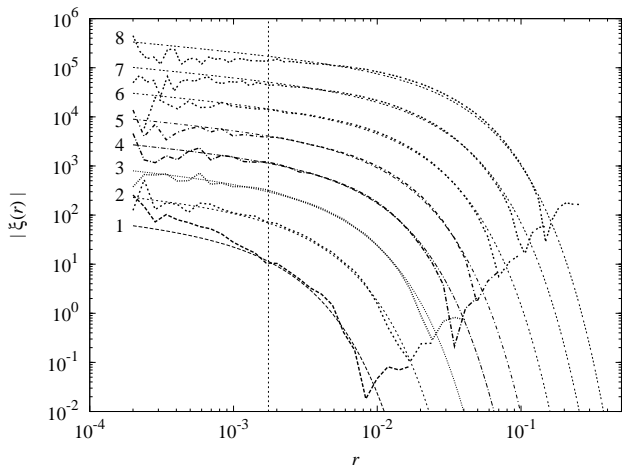}
\caption{Comparison of $\xi(r,t)$ measured in SL128 with the formula
in Eq.~\eqref{eq:scalbehav}, using the rescaling of
Eq.~\eqref{eq:rescaling}. For clarity the
amplitudes of the different curves have been rescaled (by a factor
$3^{t-1}$). Moreover for all the curves, we have plotted only the
scales such that $\xi(r,t)\gtrsim 0.1$ since Eq.~\eqref{eq:scalbehav}
does not describe smaller amplitude correlations. The vertical line
corresponds to the softening length $\varepsilon$.}
\label{fig:expcut}
\end{figure}

In Fig.~\ref{fig:Rst} is shown the evolution of the rescaling factor
$R_s(t)$ found in constructing Fig.~\ref{fig:collapse}, as a function
of time, with the (arbitrary) choice $R_s(1)=1$. Shown in Fig.~\ref{fig:Rst} 
are also three (theoretical) curves, which we will explain in
the next subsection. Before this we remark on two further aspects of
the scaling relation which are worth noting:

\begin{itemize}
\item The function $\Xi(r)$, when it is larger than
$\sim 0.1$, can be well approximated by a simple power law with
an exponential cut-off:
\begin{equation}
\Xi(r) \approx A \left( \frac{r}{\hat R}\right)^{-\gamma}
\exp\left(-\frac{r}{\hat R}\right)  \;, 
\label{eq:scalbehav} 
\end{equation} 
where we have estimated (see Fig.~\ref{fig:expcut}) the following
values for the three parameters: $A=40$, $\gamma=0.28$ and $\hat
R=1.45 \cdot 10^{-3}$. This last parameter gives the normalization of
$\xi(r,t)$ at $t=1$.  In order to see how well this fit
describes the evolution, we show in Fig.~\ref{fig:expcut} both the
data and the curves inferred from it, using $R_s(t)$ as measured.

\item  Since we have defined the homogeneity scale
$\lambda$ by $\xi(\lambda)=1$ it is clear that, once the
spatio-temporal scaling relation is valid, we have $\lambda(t) \propto
R_s(t)$.

\item Since the PS and mass variance are simply related
to $\xi(r)$, we expect the scaling relation to be reflected in one for
these quantities as well. We will see to what extent this is the case
below.

\end{itemize}

\subsubsection{Spatio-temporal scaling and ``self-similarity''}

We have observed that, from $t \sim \tdyn$, the two-point correlation
function, at least down to $\xi(r) \sim 0.01$ (level of estimator
noise) obeys to a good approximation the spatio-temporal scaling
relation Eq.~(\ref{eq:rescaling}), with the measured $R_s(t)$ shown in
Fig.~\ref{fig:Rst}. In this section we discuss this result, in
particular its relation to similar behaviors which have been
studied in cosmology.

In the context of cosmological N body simulations this kind of
behavior, {\it when} $R_s(t)$ {\it is itself a power law (in time)},
is referred to as {\it self-similarity}. Such behavior is expected in
an evolving self-gravitating system (see
e.g. \cite{peebles,efstathiou_88,bertschinger}) because of the
scale-free nature of gravity, {\it if} the expanding universe model
{\it and} the initial conditions {\it contain no characteristic
scales}. Initial conditions in N-body simulations do, however,
necessarily contain one such scale, which is associated to the particle
discreteness (i.e. the grid spacing $\ell$ in the case of a perturbed
lattice). Further, as we have discussed above, simulations introduce
(at least) two further scales: the box-side $L$ and force softening
$\varepsilon$. Thus self-similarity is expected to be observed in N
body simulations of an Einstein-de Sitter model (i.e. a flat matter
dominated universe), starting from pure power-law initial PS $P(k)
\sim k^n$, {\it if all effects associated with these length scales can
be neglected}. 

On theoretical grounds there are different expectations
(\cite{bertschinger,bertschinger2,efstathiou_88}) about the range of
exponents $n$ of the PS which should give self-similar
behavior. Effects coming from the particle discreteness are expected
to become less important as the PS becomes ``redder'' (i.e. smaller
$n$, with more relative power at larger scales), while a PS
which is too ``red'' will become sensitive to finite size effects
(i.e.  to the box size). A more quantitative analysis of the
dependence of dynamically relevant quantities (e.g. the variance of
velocity and force fields) on these ultra-violet and infra-red cut-offs
suggests that self-similarity should apply in the range $-1<n<+1$, and
such behavior has in fact been observed, to a good approximation, to
apply in simulations in an EdS universe of such spectra
\cite{efstathiou_88,smith}. 
While there has been considerable discussion also of the case
$-3<n<-1$ in the literature, with different conclusions about the
observed degree of self-similarity, the case $n \geq 1$ has remained
open
\footnote{The reason why the case $n>1$ has not
been studied numerically appears to be twofold: firstly, it is not of
direct interest to ``real'' cosmological models which typically
describe PS with exponents in the range $-3 < n < -1$; secondly, the
simulation of such initial conditions is considered ``hard to simulate'' 
(see e.g. \cite{smith}).}. In our discussion below we will see in greater
detail why the cases $n>1$ and $n<1$ are expected to be possibly very
different with respect to ``self-similarity''.

Our results above clearly suggest that what we
have observed is a simple generalization of this 
self-similarity to a static universe (in which there is evidently
also no characteristic length scale), and to the case 
$n=2$. Let us examine more carefully whether this is the
case, by generalizing to the static case the argument 
(see e.g. \cite{peebles}) used to derive the power-law 
behavior of $R_s(t)$ in an expanding universe.

In order to derive this behavior of $R_s(t)$, we 
{\it assume} that the spatio-temporal scaling relation 
holds exactly, {\it i.e., at all scales}, from, say,
a time $t_{s} > 0$. For $t > t_{s}$ we have then 
\begin{equation}
\begin{split}
P(\ve k,t) &=  \int_{L^3} \exp(-i\ve k\cdot \ve r) \,{\xi}(\ve
r,t) \, \D[3] \ve r \\
&=  R_s^3(t) \int_{L^3} \exp(-i R_s(t) \ve k\cdot \ve x ) \,
\Xi(|\ve x|) \
\D[3] \ve x   \\
&=  R_s^3(t) P(R_s(t)\ve k,t_s)  \ .
\label{eq:Rsdetermine}
\end{split}
\end{equation}
where we have chosen $R_s(t_s)=1$. Assuming now that 
the PS  at small $k$ is amplified as given by linear theory,
i.e., as in Eq.~(\ref{eq:pk_evolution_linear}), one infers
for any PS $P(k) \sim k^n$ (and $n<4$ so that linear theory
applies):
\begin{equation} 
R_s(t) = 
\left( \frac{\cosh \frac{t}{\tdyn}}{\cosh \frac{t_s}{\tdyn}} \right)
^\frac{2}{3+n} 
\xrightarrow{t\gg t_s}
\exp\left[ \frac{2(t-t_s)}{(3+n)\tdyn} \right]\,.
\label{eq:predRst}
\end{equation}
In the asymptotic behavior the relative rescaling in space 
for any two times becomes a function only 
of the {\it difference} in time between them so that we can write
\begin{equation} 
\xi (r, t + \Delta t) = \xi \left(\frac{r}{R_s(\Delta t)}, t \right)
\; ; \quad R_s(\Delta t) = e^{ \frac{2 \Delta t} {(3+n) \tdyn} } \,.
\label{eq:self-sim-xi}
\end{equation}
This is analogous to what is called self-similarity in EdS
cosmology. In that case the linear theory describes a growing and a
decaying mode, both of them power laws in time. Asymptotically
$R_s(t)$ is thus itself a simple power law \footnote{One has 
\cite{peebles,efstathiou_88} $P(k,t) \propto t^{4/3} k^n$ at small $k$, and thus
$R_s(t) \propto t^{\frac{4}{3(3+n)}}$.}.

Let us now return to Fig.\ref{fig:Rst}. In addition to the measured
values of $R_s(t)$ the figure shows two curves corresponding to
Eq.~(\ref{eq:predRst}) with $n=2$. The first (labeled ``expected'')
corresponds to taking $t_s=1$ in the derivation above, i.e., assuming
that the scaling relation holds at all scales for $t>1$. The second
(labeled ``expected (resc)'') is the same functional behavior, but
rescaled by a constant to give a good fit to the larger time (from $t>
t_s \sim 2.5$) behavior. This latter behavior is clearly very
consistent with the relation given in Eq.~(\ref{eq:predRst}): starting
from this time the slope is very close to constant and equal to
$\frac{2}{5}$ in units of $\tdyn$.

Our results are thus indeed clearly well interpreted as a
generalization in a static universe of self-similarity as observed in
simulations EdS universes, for ``redder'' spectra. This
self-similarity sets in, however, from about $t_s=2.5$, while we
observed the spatio-temporal scaling relation already to apply
approximately from $t \sim 1$ ($\approx \tdyn$). Note that the fact that the
functional behavior of $R_s(t)$ in $1< t < 2.5 $ is
inconsistent with Eq.~(\ref{eq:predRst}) with $n=2$ {\it implies} that
the spatio-temporal scaling relation cannot hold at all scales at
these times: specifically it cannot hold at small $k$, where $P(k)
\propto k^2$, as we have seen that at these scales the PS {\it is}
linearly amplified at this time.

A possible explanation for this behavior is suggested by the third
curve (labeled ``Poisson'') shown in Fig.\ref{fig:Rst}.  This curve
corresponds to Eq.~(\ref{eq:predRst}) with $n=0$ and $t_s=1$. The fact
that it fits the points reasonably well --- although not so well as
the $n=2$ theoretical curve for $t>2.5 $ --- suggests the following
interpretation: between $1< t < 2.5 $ we are observing a first phase
of ``self-similarity'', restricted to smaller scales, where the
initial PS is roughly flat (i.e. Poisson-like with $n=0$) in a small
range of $k$ around the Nyquist frequency (see
Fig.~\ref{fig:pk_SL128}).  Such an interpretation is consistent with
the fact that the $\xi(r)$ in the non-linear regime observed in
simulations from Poissonian initial conditions is, to a very good
approximation, the same as that observed from SL initial conditions
\cite{Baertschiger:2002tk,Baertschiger:2004tx}.
On the other hand, the wave modes at which the PS is Poisson-like are
very large --- of the order of the inverse of the inter-particle
spacing --- and so the observation of apparent self-similarity driven
by these fluctuations is somewhat surprising: such behavior is
expected, as we have discussed above, when the effects of discreteness
may be neglected.  We will see below that this interpretation of the
spatio-temporal scaling observed in the correlation function at
non-linear scales at early times --- as a first phase of
self-similarity driven by Poisson fluctuations at small scales --- is
not correct. In particular it is reproduced in the smaller simulations
we will analyze below in which there is no initial Poisson plateau
around $k_N$. Further we will see that the form of the non-linear
correlation function is already established at times when two-body
correlations due to nearest neighbor interactions are the dominant
source of correlation at these scales.

\subsubsection{Evolution of the mass variance}

In this section we study the normalized mass variance $\sigma^2 (r)$,
defined in Eq.~(\ref{eq:variance}). Through the study of this quantity
we can probe further the scaling properties (and self-similarity) we
have just seen. We can also explain and see the interesting and
non-trivial differences in this respect between the case of a PS with
$n<1$ and $n>1$.

\begin{figure}
\includegraphics[width=0.5\textwidth]{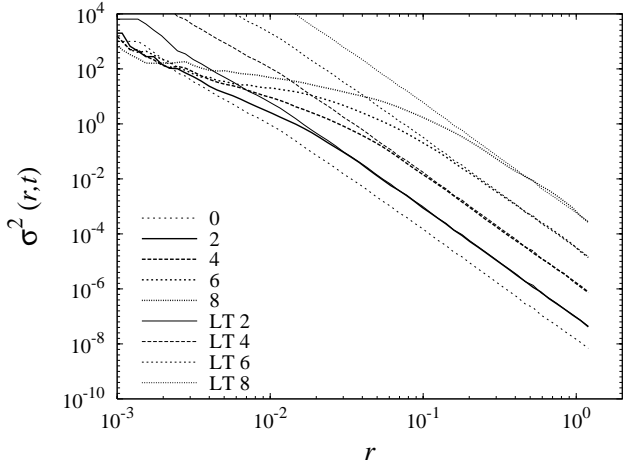}
\caption{ Evolution of the mass variance in SL128 at times $t=0,2,4,6,8$,
together with the predictions of Eq.\eqref{eq:asigma} (labeled as LT).}
\label{fig:variance}
\end{figure}

Given that the mass variance is expressible (cf. Eq.~\ref{sigma-ps})
as an integral of the PS, one might anticipate that it 
will show, at large $r$, the same behavior as  the PS at small 
$k$, i.e., we expect to find the simple scale independent 
amplification of linear theory:
\begin{equation}
\sigma^2(r,t)= A_{\sigma^2}(t) \sigma^2(r,0)
\label{linear-amplification-sigma}
\end{equation}
where
\begin{equation}
\label{eq:asigma2} 
A_{\sigma^2}(t)=\cosh^{2} (t/ \tdyn) \propto R^{3+n}_s(t) \;  
\end{equation}
for a PS $P(k) \propto k^n$ at small $k$.

\begin{figure}
\includegraphics[width=0.5\textwidth]{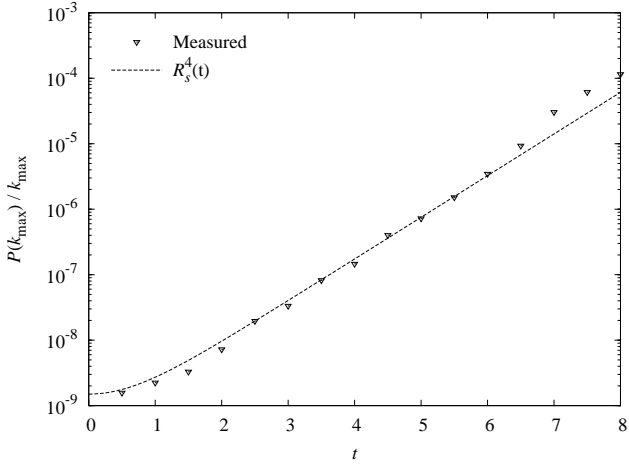}
\caption{Behavior of $P(k_\text{max},t)/k_\text{max}$ as a function of
time measured in SL128. The dashed line represents the
behavior given in Eq.~(\ref{kmax-evolution}). }
\label{fig:kmaxPmax}
\end{figure}

In Fig.~\ref{fig:variance} is shown the temporal evolution in SL128 of
$\sigma^2(r)$. At each time we observe at large $r$ the behavior
$\sigma^2(r) \propto 1/r^4$ characteristic of a SL. The dotted lines
show the best fit to the behavior of Eq.~(\ref{linear-amplification-sigma})
above, which we find is 
\begin{equation}
\label{eq:asigma} 
A_{\sigma^2}(t)=\cosh^{8/5} (t/ \tdyn) \propto R^4_s(t) \;.
\end{equation}
rather than the anticipated behavior of Eq.~(\ref{eq:asigma2})
for $n=2$.

The reason for this discrepancy is, as we now discuss, very simple.
It is of importance as it is makes explicit the difference 
between the cases of PS with $n<1$ and $n>1$. Indeed examining
the integral Eq.~(\ref{sigma-ps}) in closer detail it turns
out that there is a qualitatively different behavior in the two
cases. For $-3<n<+1$ the integral is dominated by modes $k \sim
r^{-1}$ and one has
\be
\sigma^2(r,t) \approx C k^3 P(k,t) 
\label{sigma-n<1}
\ee
where $k=1/r$ and $C$ is a constant pre-factor which depends on
$n$. From this it follows that linear amplification of the PS at small
$k$ gives linear amplification of the mass variance at large
scales. For $n>1$, however, the integral in Eq.~(\ref{sigma-ps}) with
$P(k) \sim k^n$ at all $k$ diverges, and an ultraviolet cut-off
$k_c$ above which $P(k)/k$ decays to zero is required to regulate it
\footnote{Such a
cut-off necessarily exists in any particle distribution as $P(k)$ cannot
diverge for large $k$. One necessarily has, as we have discussed, $P(k)
\rightarrow 1/n_0$ for $k\rightarrow \infty$ \cite{book}.}.   
The effect of the cut-off is to give
\[ \sigma^2(r) \sim k_c^{-1}P(k_c)/r^4 \;,\] 
at sufficiently large $r$. 
Thus, for $n>1$ the evolution of the mass variance at large $r$ (and
thus at small amplitude) is sensitive to the evolution of the cut-off
in the PS (and the amplitude of the PS at this cut-off). From
Fig.\ref{fig:pk_SL128} we expect that in our system the role of $k_c$
will be played by $k_{\rm max}(t)$, the wavenumber at which the PS
reaches its maximum, and so we will have, at large $r$
\be
\sigma^2(r,t) \sim \frac{k_{\rm max}^{-1}(t) P(k_{\rm max},t)}{r^4} \,.
\label{evolution-sigma}
\ee
From Fig.\ref{fig:pk_SL128} we see that $k_\text{max}$ is clearly in
the range in which the amplification in $k$ space is non-linear. Thus
the evolution of this quantity, even at very large scales, is
determined by modes in $k$ space which are in the non-linear regime.
Given the time evolution we have observed for $\sigma^2(r,t)$ in
Fig.\ref{fig:variance}, we must have
\be
k_{\rm max}^{-1} (t) P(k_{\rm max},t)
\sim R_s^4(t) \,.
\label{kmax-evolution}
\ee
In Fig. \ref{fig:kmaxPmax} we see that this behavior is indeed well
approximated. It is in fact evidently the direct
consequence of the self-similarity as it is reflected in the variance,
and, equivalently, in $k$ space. To the extent that both quantities
approximate the self-similarity observed in $\xi(r,t)$, any length
scale derived from either the variance or PS must scale $\propto
R_s(t)$.  Thus, in particular, $k^{-1}_{\rm max} \propto R_s(t)$ and
the maximum value of the PS, which has dimensions of volume, must scale as
$P(k_{\rm max},t) \propto R_s^3 (t)$ in this regime.

\begin{figure}
\includegraphics[width=0.5\textwidth]{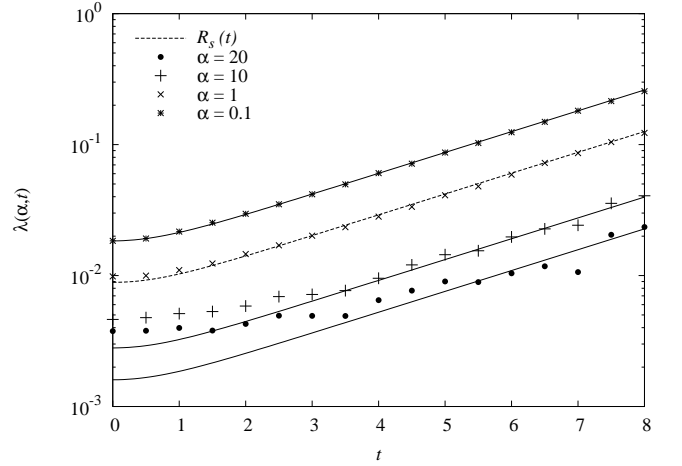}
\caption{Behavior of $\lambda(\alpha,t)$, the length for which 
$\sigma^2(r,t)=\alpha$ for $\alpha=0.1,1.0,10.0,20.0$ in the 
simulations SL128.}
\label{fig:homo-scale-variance}
\end{figure}

It is instructive also to examine a little further how the
spatio-temporal scaling behavior, and self-similarity, are
approximated in the variance. In order to illustrate this 
we consider the
temporal evolution of scales $\lambda(\alpha,t)$ defined by the
relation 
\be
\sigma^2\left(\lambda(\alpha,t),t\right)=\alpha
\ee
 where $\alpha$ is a chosen constant. If there is a spatio-temporal
 scaling in the system we should find that $\lambda(\alpha,t) \propto
 R_s(t)$.  In particular any choice $\alpha
\sim 1$ gives, as we discussed in Sec.\ref{sec2}, a reasonable
definition for the homogeneity scale, which should be equivalent to
the one we have taken above ($\xi(\lambda,t) = 1$ once non-linear
clustering has developed). In Fig. \ref{fig:homo-scale-variance} we
show $\lambda(\alpha,t)$ for $\alpha =20,10,1,0.1$; also shown are
curves proportional to $R_s(t)$ in the self-similar regime (i.e. as
given by Eq.~(\ref{eq:predRst}) with $n=2$). The figure illustrates
nicely how the scaling applies only at large scales (corresponding to
smaller fluctuations) initially and then propagates to smaller (more
non-linear) scales. At the time $t \approx 2.5$, which we 
identified above in our analysis of $\xi(r,t)$ as the time from which 
self-similarity is well approximated, the scaling behavior given
by $R_s(t)$ is manifestly well approximated well for $\alpha \gg 1$,
i.e., into the non-linear regime.  We do not, however, see a behavior
consistent with the hypothesis that the evolution prior to this time
($1 < t < 2.5$) is self-similar and associated to a PS with $n=0$
around $k_N$: this would correspond as in Fig.\ref{fig:Rst} to a
faster evolution of the scales shown here at these times, which is not
what is observed.


\subsubsection{Self-similarity and the regime of validity of linear theory}

\begin{figure}
\includegraphics[width=0.5\textwidth]{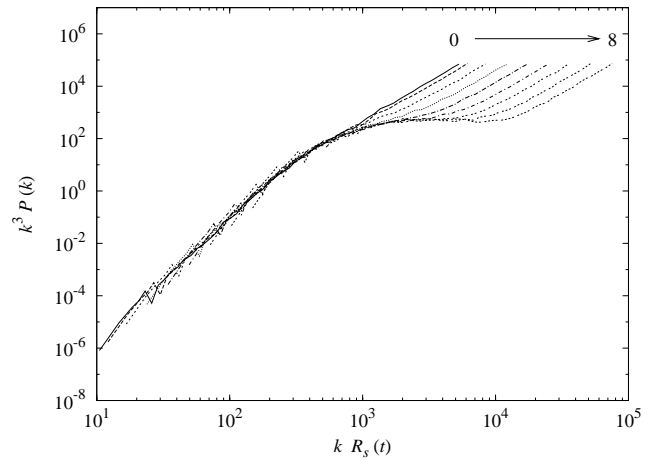}
\caption{Collapse plot 
of $\Delta^2(k,t)$ for the SL128 simulation using as scaling factor
$R_s(t)$ as described in Eq.\eqref{eq:sf1}.}
\label{fig:kcubePS-rescaled}
\end{figure}

The derivation of $R_s(t)$ in Eq.~(\ref{eq:predRst}) explains
implicitly the physical origin of the self-similar behavior: 
if the small $k$ PS is a simple power law, the evolution
of the two-point correlation function is self-similar, with
$R_s(t)$ given by Eq.~(\ref{eq:predRst}), in
the approximation that fluctuations grow as described 
by the linearised fluid theory.   
Self-similarity applies to the full evolution to the extent that this
self-similar temporal evolution at linear scales becomes ``imprinted''
on smaller non-linear scales. The mechanism by which this happens is
simply the collapse of the initial mass fluctuations at large scales,
at time scales fixed by linear theory.  Self-similarity is thus a good
approximation to the extent that the clustering amplitudes at any
scale depend only on the prior history of larger scales. 
In terms of power transfer in the evolution of clustering,
self-similarity can thus be interpreted qualitatively as
indicating that there is a much more efficient transfer of power 
from large to small scales than in the opposite direction. Our 
results here show that this is true also in more
``blue'' initial conditions with a small $k$ PS $P(k) \propto k^n$ and
$n>1$, in which the variance of mass in real space spheres is 
dominated by fluctuations at much smaller scales which evolve in the
non-linear regime.

These points are further illustrated in
Fig.~\ref{fig:kcubePS-rescaled}, which shows a ``collapse plot'' for
the temporal evolution of $\Delta^2 (k,t) \equiv k^3 P(k)$.  It
follows from Eq.~(\ref{eq:Rsdetermine}) that, when self-similarity
applies, we have the behavior
\be
\label{eq:sf1}
\Delta^2 (k,t) = \Delta^2 ( R_s(t) k, t_s) 
\ee
where, as above, $t_s<t$ is an arbitrary initial time and $R_s(t)$ is
given by Eq.~(\ref{eq:predRst}). In Fig.~\ref{fig:kcubePS-rescaled} is
plotted the rescaled function at each time, starting from
$t_s=0$. At small $k$, we see that right from the initial time 
the self-similarity is indeed followed (as the rescaled curves 
are always superimposed at these scales). This is simply because 
linear theory, which is valid at these scales, gives such a
behavior. As time progresses we see the range of $k$ in which
the curves are superimposed increases, extending into the
non-linear regime. 
Thus the
self-similarity ``propagates'' progressively from small $k$ to larger
$k$, carried by the scales which are evolving non-linearly. Note that the
behavior at asymptotically large $k$ is simply $\Delta^2 (k,t)
\propto k^3/n_0$ (where $n_0$ is the mean particle
density) at all times, corresponding to the shot noise present in all
particle distributions with average density $n_0$ and which by
definition does not evolve in time\footnote{Note that in the two-point correlation function
this time independent discrete contribution appears as a singularity
at the origin. It therefore does not ``pollute'' the collapse plots
for $\xi(r,t)$. This also explains why one is able to identify the
scaling behavior more readily by eye in this quantity. The collapse
plot for $\sigma^2(r,t)$, which we have not shown, is similar to that
for $\Delta^2(k)$.}.

\begin{figure}
\includegraphics[width=0.5\textwidth]{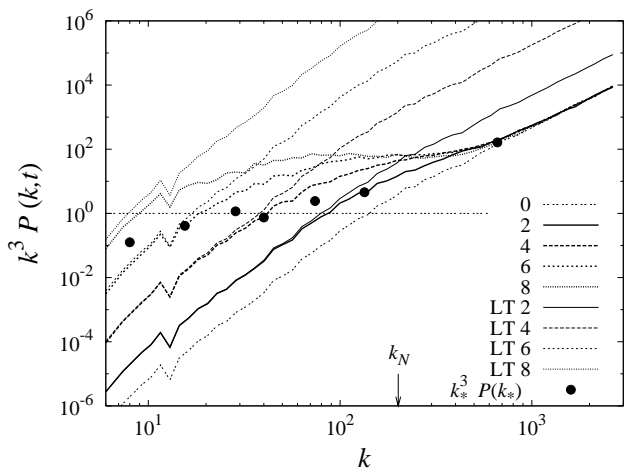}
\caption{Behavior of $\Delta^2(k,t)$ in SL128 together
with the prediction of linear theory (LT). The points correspond to
the value of $\Delta^2(k_*,t)$, at times (from right to left) 1, 2, 3,
4, 5, 6 and 7, where $k_*$ is the wave-number above which the
evolution the PS is no more longer well approximated by linear theory. 
\label{fig:kcubePS}}
\end{figure}

Let us finally return to the question of the breakdown of linear
theory. In our discussion of Fig.~\ref{fig:pk_SL128} in
Sec.~\ref{ps-evolution} above, we noted that the scale-independent
amplification of linear theory describes very well the evolution
of the PS up to a wavenumber $k$, which we denoted $k_*(t)$
and which decreases with time. A question of interest is what the 
criterion is  which determines this scale, i.e., what the criterion 
is for the application of linear theory. We cannot answer this
question rigorously without considering, at least, the next order
in this perturbative treatment \footnote{It is in fact
possible \cite{peebles} to write the equation for
the evolution of density fluctuations in $k$ space in
a convenient form for this purpose, with  all corrections
to the linearised fluid limit in two formally simple terms.
See also \cite{Padmanabhan:1989gm} for discussion of these issues.}.
We will not attempt to do so here, but rather consider
determining such a criterion phenomenologically 
(i.e. from the simulations).  

In principle this criterion may, in general, be quite 
complicated, as it would be expected to depend on the 
fluctuations present at all scales. Once we are in the 
self-similar regime, however, we expect that all characteristic
scales in $k$ space, and in particular  $k_*(t)$, should 
scale $\propto R_{s}^{-1} (t)$. Such a time dependence results 
if one supposes $k_*(t)$ determined by a dimensionless 
quantity having some given amplitude. The evident simple
criterion which suggests itself is
\be
\label{criterion-LT}
\Delta^2 (k_*(t), t) = \text{constant}\,.
\ee
Fig.~\ref{fig:kcubePS} shows the evolution of $\Delta^2 (k, t)$,
together with the evolution in linear theory. The points 
(small black circles) mark the approximate value of $k_*$ at 
each time, determined as the scale at which the full evolution
deviates from the linear theory in each case\footnote{The value of $k_*$ in
Fig.~\ref{fig:kcubePS} (used for the points) has been estimated using the
following criterion: $| \ln P(k_*,t) / \ln P_\text{LT}(k_*,t) - 1 | =
0.05$, where $P_\text{LT}(k,t)$ is simply the initial PS amplified by
linear theory, i.e., Eq.~\eqref{eq:pk_evolution_linear}. }.
The horizontal line shows that, starting from about $t=3$, when
the self-similarity has set in, Eq.~(\ref{criterion-LT}) with
the constant set equal to unity is a reasonably good fit 
to the observed $k_*(t)$. The deviation of the last point, at $t=7$
can be attributed to finite size effects, as we see that at this
time the smallest $k$ modes in the box are no longer described
well by the linear evolution.

For $n<1$, because of Eq.~(\ref{sigma-n<1}), the criterion 
Eq.~(\ref{criterion-LT}) for the breakdown of linear theory
is equivalent to one stating it as a threshold value of the 
real space variance. In the current case, with $n>1$, there
is no such evident equivalence, as the mass variance at
scales $ r > k_*^{-1}(t)$ are not directly determined
by the fluctuation of these modes, but rather by the
fluctuations in larger $k$ modes. Thus in this case the 
physical criterion for the breakdown of linear theory is
really more appropriately given in $k$ space
\footnote{If one wishes to define a real-space scale directly, this 
can be done by using the mass
variance defined in a Gaussian window, i.e., with $\tilde W(k;r)$ in
Eq.\eqref{sigma-ps} given by a Gaussian of width $1/r$. This is really
just a trivial way of restating the $k$ space condition in real
space.}.

\subsubsection{Role of two-body correlations}

The gravitational force on a particle in an SL is dominated, for 
small $\delta$, by that exerted by its six nearest neighbors, and 
for large $\delta$, by its single nearest neighbor \cite{gabrielli_06}. 
One thus expects that, at sufficiently early times, the dynamical
evolution should be well approximated by neglecting all but these
dominant contributions to the force.  It has in fact been shown in
\cite{Baertschiger:2004tx} that the early time evolution of
simulations of small $\delta$ SL can be well approximated by a two
phase model: in a first phase the particle moves under the effect of
its six nearest neighbors, and then subsequently, when the lattice
symmetry is broken, under the effect only of a single nearest
neighbor. The first non-linear correlations then emerge as these
nearest neighbor pairs fall toward one another.

\begin{figure}
\includegraphics[width=0.5\textwidth]{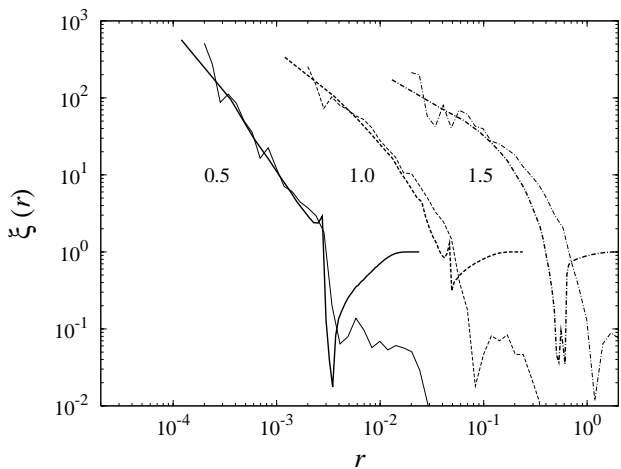}
\caption{Two-point correlation function in SL128 at times $t=0.5,1,1.5$ 
(thin lines) together with the approximation got from the NN PDF
(thick lines). For clarity the behaviors at different times 
have been arbitrarily rescaled on the $x$-axis.
\label{gamma_SL128_nn}}
\end{figure}

As described in Sec.\ref{sec2} the relation Eq.\eqref{eq:omega1} holds
in the approximation that the correlations are primarily due to
correlated pairs of nearest neighbor particles. Its validity is thus a
good probe of the probable adequacy of a dynamical model like that
just described.  In Fig.~\ref{gamma_SL128_nn} is shown the comparison
of the two relevant quantities: the correlation function measured in
the simulation and the one reconstructed by using
Eq.\eqref{eq:omega1}, i.e., by considering explicitly only nearest
neighbors correlations. We see that the relation holds very well until
$t=1$. 

This is a quite striking and surprising result: the form of
$\xi(r,t)$ --- which is subsequently that which is observed to scale in
the asymptotic self-similar evolution --- has already emerged at a time
when nearest neighbor interactions play a crucial role in the
dynamics. In the previous sections, however,  we have seen that 
this asymptotic behavior is characterised by a time dependence
derived in a fluid limit. Such a limit is normally expected to
be valid in the opposite case that two or few body interaction 
with nearest neighbor particles can be neglected (rather than
being dominant in the approximation just considered). Indeed
we noted that when the asymptotic scaling behavior there is
necessarily no explicit dependence on the characteristic length 
scales in the system ---  and notably those associated with 
the discreteness of the distribution which directly enter in
determining the strength of nearest neignbour forces.
We will discuss this point further below after a presentation of
results of the other SL simulations we have performed.

\subsubsection{Dependence on the normalized shuffling parameter }

As discussed above SL initial conditions may be characterized, for
their gravitational evolution, by the single dimensionless parameter
$\delta$. Our analysis until now has concerned solely the simulation
SL128 and thus only a single value ($\delta=1$). Our primary result
--- that this system tends in a few dynamical times to
a``self-similar'' evolution --- would be expected to be true for any
(finite) value of $\delta$: this particular spatio-temporal scaling
behavior is determined solely by the $k^2$ {\it form} of the small $k$
PS, which is invariant under changes in $\delta$.  Thus the only thing
that we would expect to change non-trivially when $\delta$ changes is
the transient regime to the asymptotic self-similar
behavior. Specifically we might expect both the duration of this
transient and its nature to change.

The emergence of self-similarity in the evolution corresponds, as we
have discussed at length, to a behavior which is explicitly
independent of the  discreteness scale $\ell$ characterizing its 
particle-like nature. The simplest interpretation of this behavior
--- and the usual one in cosmology --- is that this corresponds
to a fluid-like behavior of the system i.e. to an evolution which
can be described,  at both linear and non-linear scales, by a set 
of non-linear fluid equations approximating the particle dynamics
\footnote{More
precisely the system is assumed then to evolve as described
by a set of Vlasov-Poisson equations. See appendices below.}. 
If this interpretation is correct, any $\delta$-dependent effects 
in the evolution of SL with different $\delta$, but with 
the {\it same} large scale fluctuations  (i.e. small $k$ PS) can 
then be considered as ``discreteness effects''. 

This equivalence of the fluid limit of the evolution from
SL with different $\delta$ 
can be seen even more explicitly as follows, for the case 
that $\delta$ is small. In this limit the so-called Zeldovich 
approximation to the fluid limit evolution (see Appendix \ref{app-b} below) 
is valid. Each element of the fluid then moves
according to
\be
{\bf x} ({\bf q}, t) ={\bf q}  + f(t) {\bf u} ({\bf q},t=0) 
\label{Zeldovich}
\ee
where ${\bf q}$ is a Lagrangian (time-independent) coordinate, which
we can take here to be the lattice point from which the particle/fluid
element is displaced, and ${\bf u}({\bf q},t=0)$ is the displacement 
of the particle/fluid element at the initial time. The function $f(t)$ 
is simply the growth factor of the fluctuations in linear theory.  
The effect of the evolution, in this limit, is thus manifestly
to transform  one SL into another one with a different (larger) $\delta$. 
Thus in the linear
fluid limit, starting from a small $\delta$, the evolution of the
system should be {\it identical} (statistically, and up to an overall
scale transformation) to that of an SL with a larger $\delta$.

The simulations SL64, SL32, SL24, and SL16, as we have defined them
allow us to explore the $\delta$ dependence (and thus non-fluid
effects) in the evolution from SL initial conditions.  As described
above in Sec.\ref{sec2},  we have chosen in each case a combination of
$\delta$ and $\ell$ which leaves the amplitude of the PS constant (in
the length units we have chosen, fixed by the box size in these simulations). 
This choice means that the
evolution of any two simulations in time should agree (without any
rescaling) if the evolution of both may be well described by the fluid
limit: this is governed, as we have seen, by the evolution of the
fluctuations at large scales which are identical.

These statements are of course true in the approximation that effects
introduced by the finite box-size of the simulation, the softening of
the force and any other effects of the numerical discretization of the
evolution, are negligible. We noted that in a {\it finite} box, and
with {\it softened} gravity, one has {\it two} additional parameters,
which one can choose as $\ell/L$ and $\varepsilon/L$. The simulations
SL64, SL32, SL24 and SL16 in Table~\ref{tab:sl_summary} correspond, as
we have described, to chosen {\it fixed} values of these two
parameters. In order to control for dependence on the box size $L$, 
we have chosen SL64 to have the same $\delta$ as SL128, so that the two
sets of initial conditions differ only in the box size. Thus 
these two simulations should give precisely the same
(averaged) results as long as finite size effects play no significant
role. We will not report in this paper the sensitivity of results to
the choice of $\varepsilon$.  We have, however, verified that, for a
considerable range of variation of $\varepsilon$ to smaller values
than the one used in the simulations we report here, there is no notable
effect on our results in the range of scales $r > \varepsilon$
where we assume them to be valid.

\begin{figure*}
\includegraphics[width=0.45\textwidth]{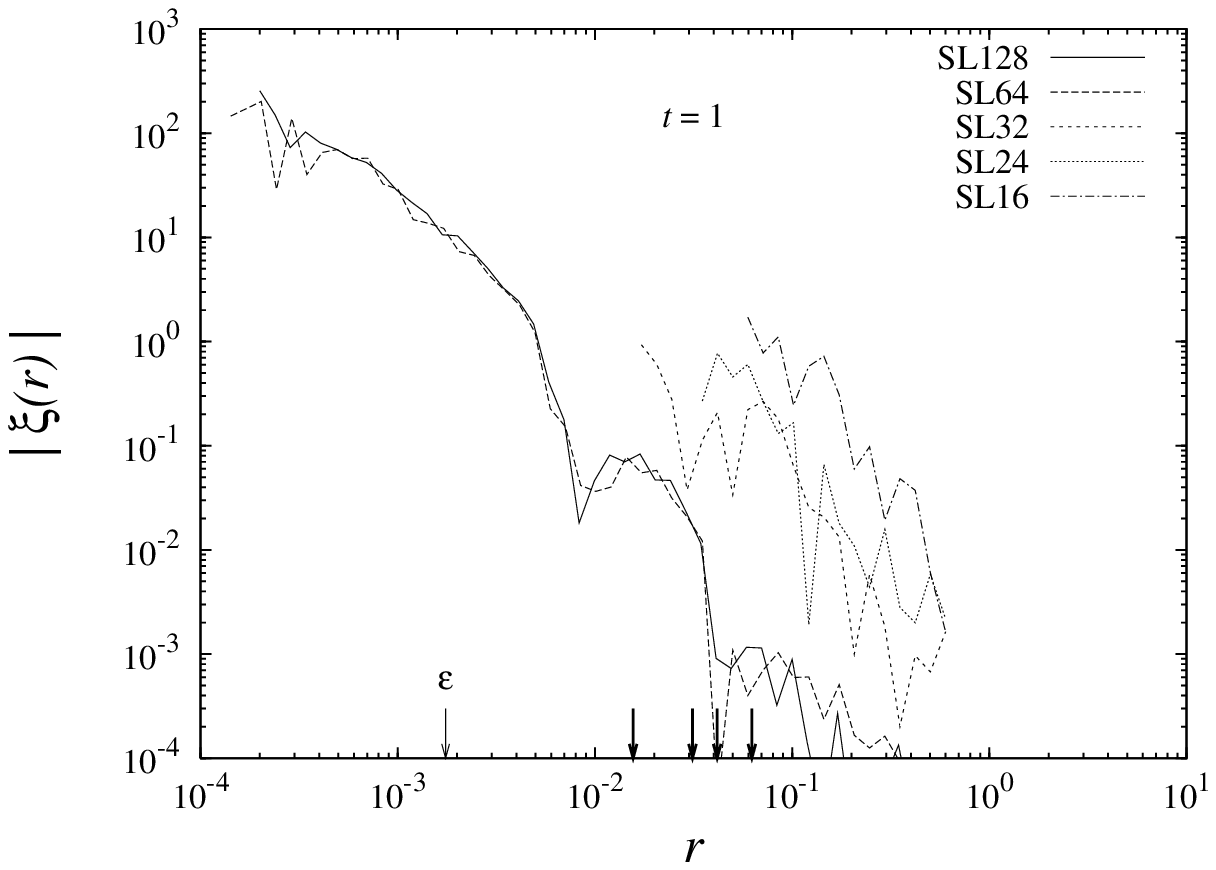}
\hspace{0.05\textwidth}
\includegraphics[width=0.45\textwidth]{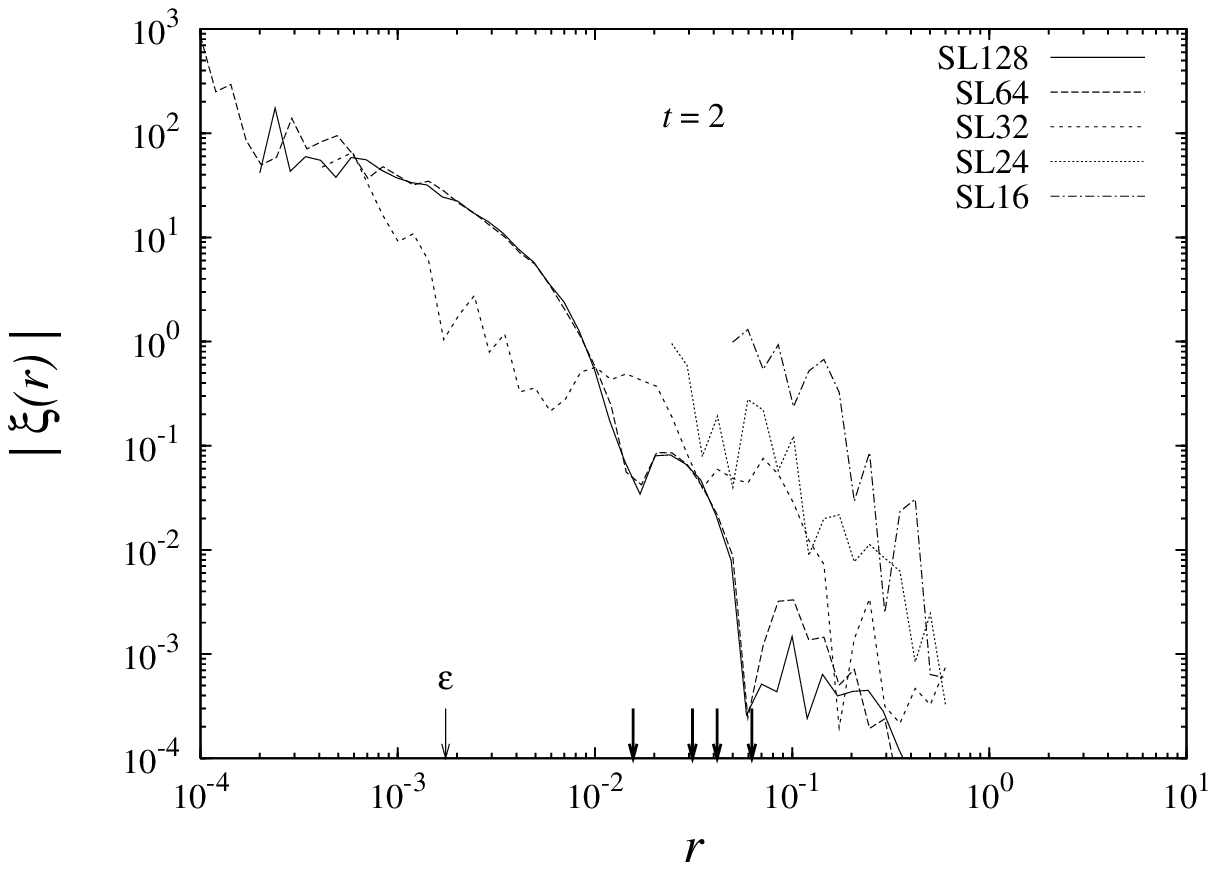}
\includegraphics[width=0.45\textwidth]{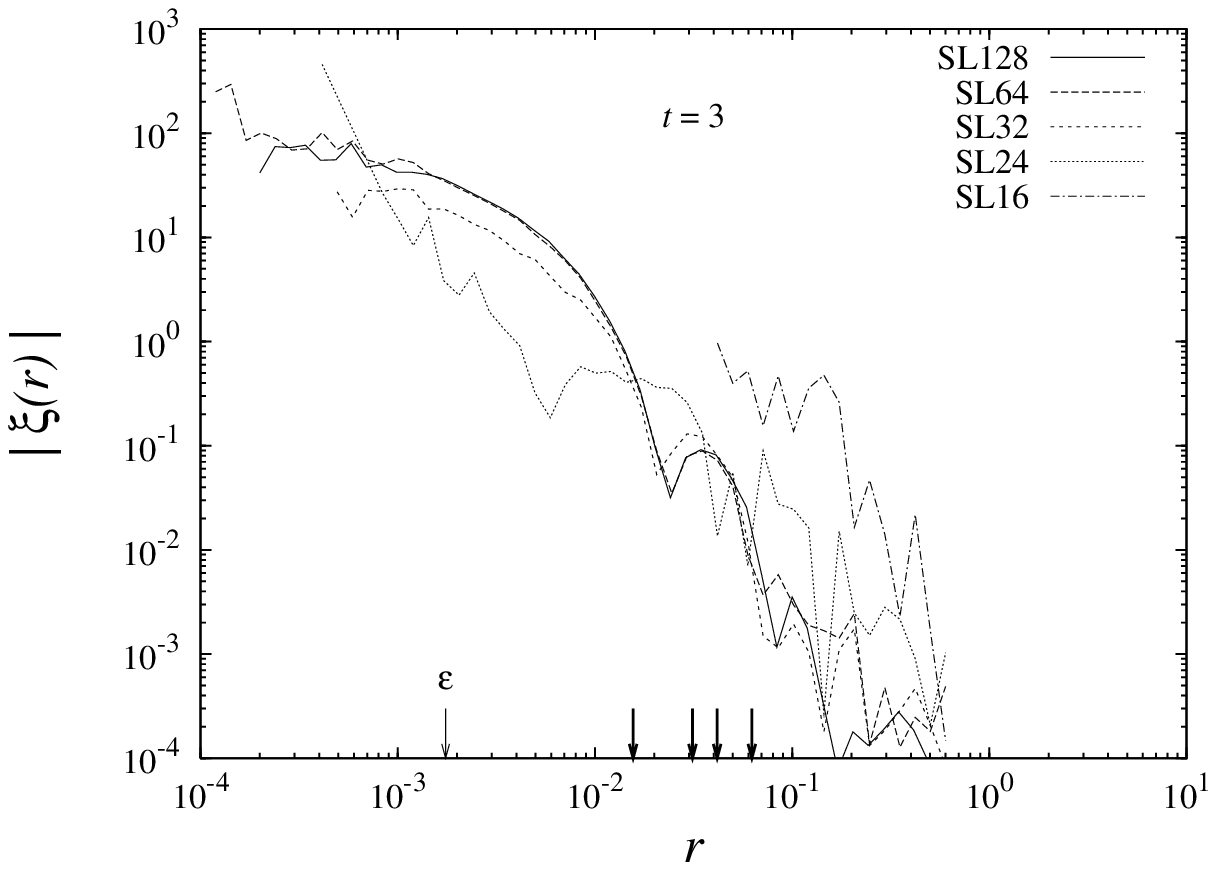}
\hspace{0.05\textwidth}
\includegraphics[width=0.45\textwidth]{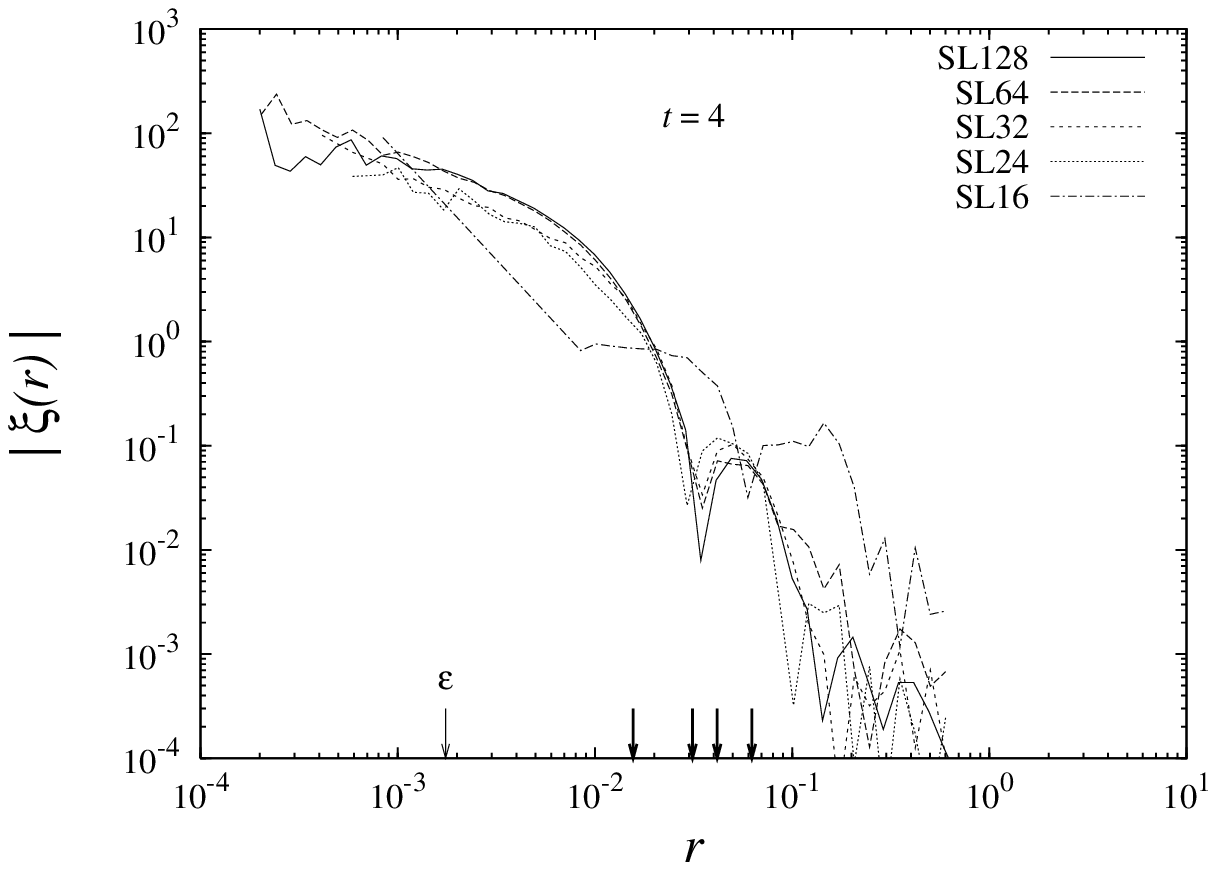}
\includegraphics[width=0.45\textwidth]{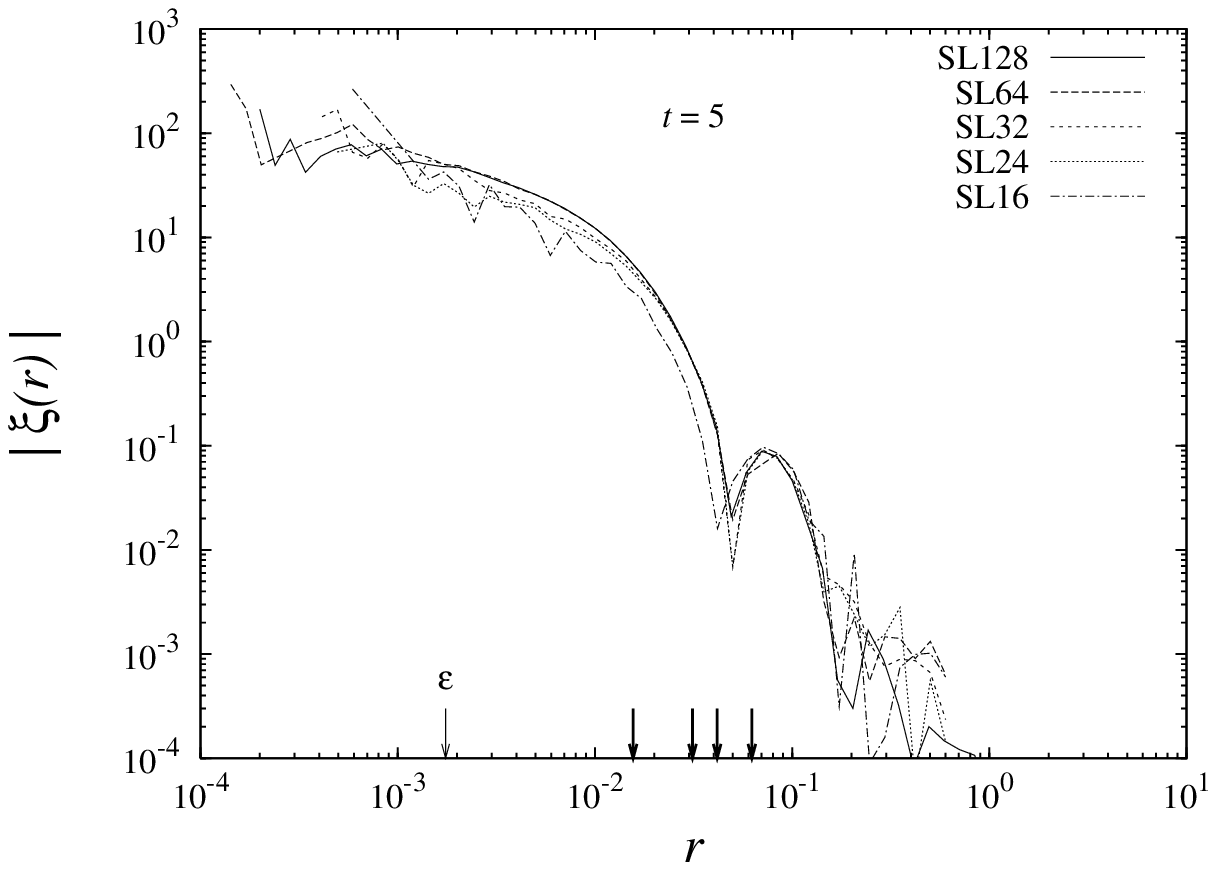}
\hspace{0.05\textwidth}
\includegraphics[width=0.45\textwidth]{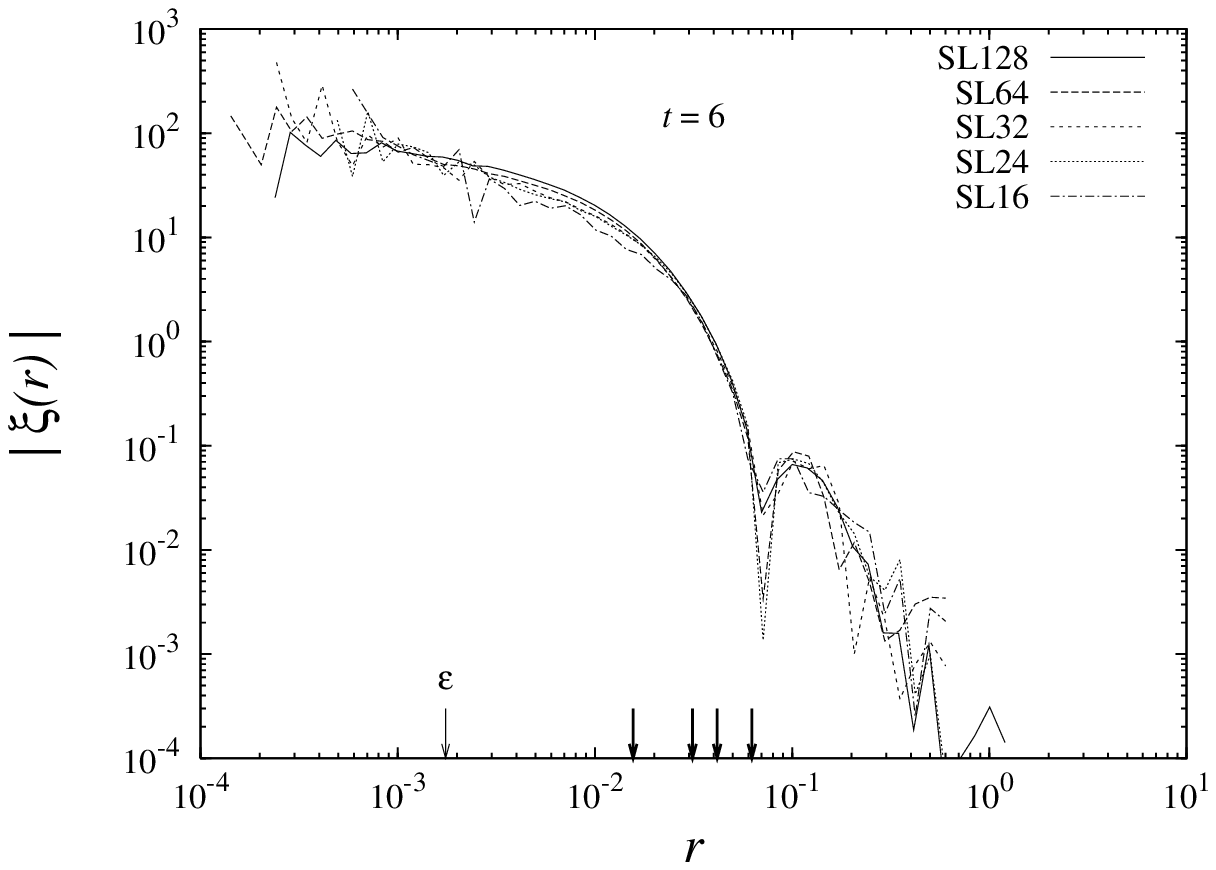}
\caption{Evolution of the two-point correlation function in the
  different simulations at times $1,2,3,4,5$ and $6$. The four
  thick arrows represent the different mesh sizes $\ell$ while the
  thin one corresponds to the value of the softening length
  $\varepsilon$. \label{fig:evol_gamma}}
\end{figure*}

In Fig.~\ref{fig:evol_gamma} are shown $\xi(r,t)$ in each of the five
SL simulations given in Table~\ref{tab:sl_summary}, for different
times, starting from $t=1$ when the structures first develop in the
largest simulations until $t=6$ when the scale of homogeneity has
reached a significant fraction of the box size.

\begin{figure}
\includegraphics[width=0.5\textwidth]{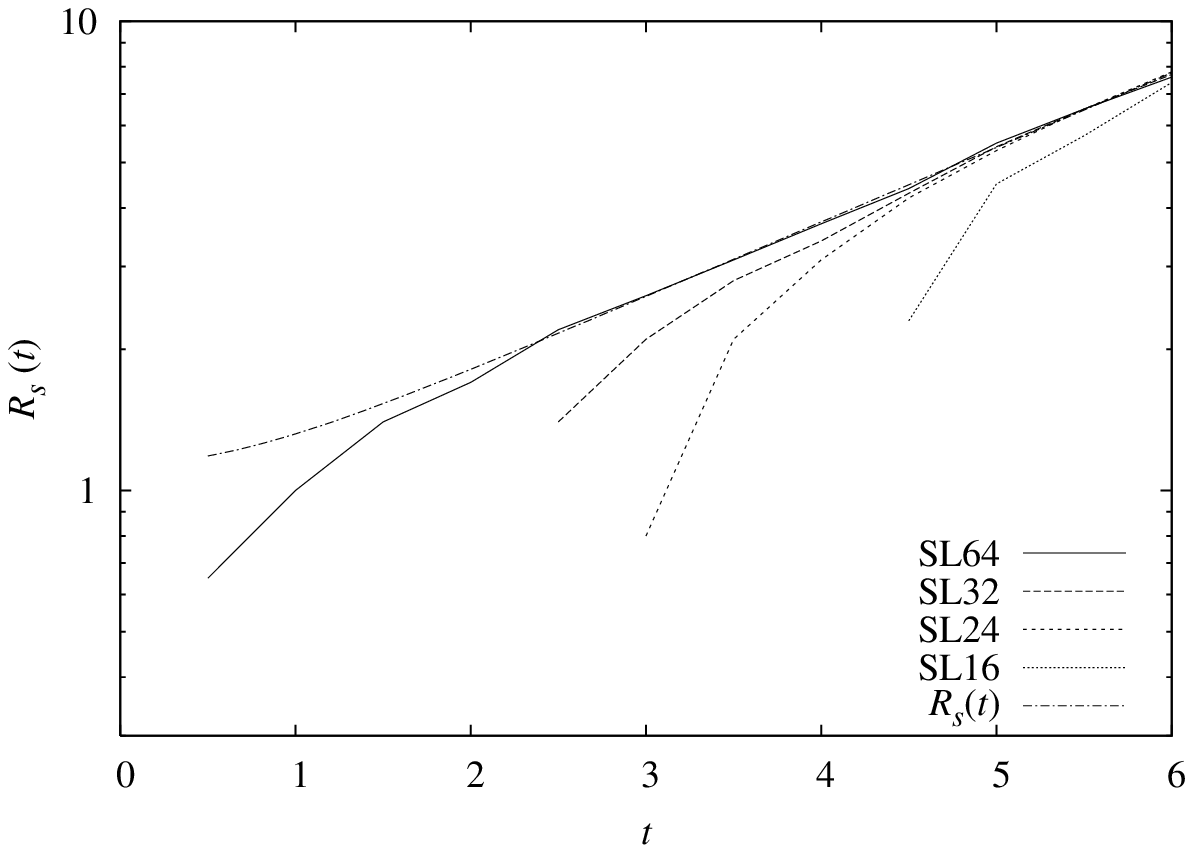}
\caption{Evolution of the rescaling factor $R_s(t)$ in the different
  simulations. Also shown is the self-similar behavior
  Eq.~\eqref{eq:predRst}.
\label{fig:lambdat} }
\end{figure}

The first point to note is the excellent agreement between the results
of SL128 and SL64, which differ only in the size of the simulation
box. This assures us that the finite size effects due to the different
values of $L$, up to the time we have shown, are very small in the
SL64 simulation, and we will assume the same is true for the SL32,
SL24, SL16 simulations (in ascribing the differences between them
solely to the change in $\delta$ and not to that in the number of
particles).

Considering now the evolution of the other simulations
we observe that:

\begin{itemize}

\item As $\delta$ decreases the time increases at which 
the system begins to evolve and form strong non-linear correlations
(i.e. develop a region with $\xi(r,t) \gg 1$). This is a qualitative
behavior expected also in the fluid limit: in the Zeldovich
approximation Eq.\~(\ref{Zeldovich}) the displacements grow
at a rate given by the function $f(t)$ which is independant
of scale. Thus, starting from a smaller $\delta$, the time at
which non-linear structures form (when $\delta \sim 1$) is 
necessarily longer\footnote{Equivalently one can say that the larger $l$
system is ``missing input power''  above its
Nyquist frequency compared to the smaller $l$ simulation.}.

\item 
When the first non-linear correlations develop there is a manifest
$\delta$ dependence in the correlation functions, i.e., the
correlations are not (statistically) equivalent to those in the larger
$\delta$ simulations. This means that at the time these correlations
emerge, the smaller $\delta$ system is not evolving as in its fluid
limit. If it were it would be in agreement with the larger $\delta$
simulation with the same initial power at the relevant scales.

\item 
Initially the non-linear correlations formed in each system ``lag
behind'' those in the larger $\delta$ simulation, i.e., $\xi(r,t)$
typically has a smaller amplitude in the smaller $\delta$ simulations.
As it evolves the smaller $\delta$ system eventually ``catches up''
with the larger ones, its correlations eventually agreeing very well
with those in all the larger $\delta$ systems over a significant range
of scale.

\item The {\it form} of the non-linear correlation function in
the asymptotic regime --- the self-similar regime we have discussed
above --- emerges to a very good approximation at a time when there is
still a quite visible ``lag'' in amplitude.

\end{itemize}

In Fig.~\ref{fig:lambdat} we show also the evolution of $R_s(t)$,
inferred in each case, as in our analysis of SL128 above, by the 
determination of the factor which describes the spatio-temporal 
scaling once it emerges as a good approximation.  We observe
that in each case we have, as for SL128, a regime of approximate 
spatio-temporal scaling of the non-linear correlation function {\it before} 
the asymptotic self-similar regime is reached. In this regime $R_s(t)$ is
smaller in amplitude than in the asymptotic regime, corresponding to
the ``lag'' of the smaller $\delta$ simulations described. However
$R_s(t)$ evolves more rapidly than in the asymptotic regime, allowing
each system to ``catch up'' with the $\delta$-independent
behavior. Note that these observations again confirm that the corresponding
regime in $R_s(t)$ in SL128 should indeed not be ascribed to a first
self-similar phase driven by the Poisson fluctuations present in this
case.

Both this ``lagging'' and the role of nearest neighbor interactions in
the formation of the first structures can be explained in the
framework of a refined version of the ``two phase model'' of
\cite{Baertschiger:2004tx} for the early time correlations. We will
present this model in detail elsewhere, and restrict ourselves here to
a few qualitative comments.

A very good approximation to the evolution of a perturbed lattice is
provided by a perturbative treatment described in
\cite{joyce_05,marcos_06}. The force acting on particles is written as
an expansion in the relative displacement of particles, in a manner
completely analogous to a standard technique used in solid state physics to
treat perturbations to crystals. One can then do a linear mode
analysis in $k$ space to determine the eigenmodes of the displacement
fields under gravity. While at small $k$ one recovers the simple $k$
independent amplification of linear fluid theory, the effect at larger $k$
(i.e. $k \sim k_N$) is, for all but some very specific modes, to slow
down the growth of fluctuations.  Thus the ``collapse time'' for
fluctuations at scales of order the inter-particle distance are indeed
slowed down, as observed here, compared to linear fluid theory.

This approximation to the early time evolution breaks down when the
force on a particle starts to be dominated by a single nearest
neighbor. At this point particles start to accelerate toward their
neighbor, giving rise to strong two-body correlations which are, as we have
seen above, the dominant contribution to the measured two-point
correlations at non-linear scales at early times. We have remarked
that, given this manifestly ``non-fluid'' mechanism for the formation
of these correlations, it is somewhat surprising to see approximately
the same two-point correlations maintained in the ``self-similar''
regime, if this regime is interpreted as the result of a purely
fluid-like evolution. Two possible, but very different, explanations
for this are the following:

\begin{itemize}

\item 
The self-similar evolution of the system in the non-linear regime is
not correctly interpreted as a manifestation of a purely fluid limit
of the N body system.  Its timescales are dictated by the fluid limit
(giving the collapse time for fluctuations at large scales), but its
non-linear dynamics are intrinsically discrete;

\item 
The early time non-linear  correlations, well described by  a discrete
dynamics,    approximate well those in    the  fluid limit because the
non-linear fluid dynamics is in fact physically well approximated by a
discrete system,  i.e., the  non-linear evolution of  the fluid,  in the
relevant phase of moderately strong non-linear correlations ($\xi(r) <
10^2$), is  well described  as the evolution    of ``lumps'' of  fluid
toward ``nearest neighbor lumps''.

\end{itemize}

We will evaluate these two quite different interpretations of our
results more quantitatively in future work.

\section{Discussion and conclusions}
\label{sec4} 

To conclude we first summarize our conclusions, and then make a few
remarks on open questions to be explored in further works.

We have studied the evolution under their Newtonian self-gravity, in
a static euclidean space, of classical point particles initially distributed
in {\it infinite} space in a quasi-uniform manner. This is a paradigmatic
problem of the out of equilibrium statistical mechanics of long range
interacting systems, which has received little attention in this context. 
Specifically we have considered a one relevant parameter 
class of initial conditions 
in which the particles are randomly perturbed off a lattice. We have 
found that our simulations converge aysmptotically (but for times 
smaller than those at which the size of the finite simulation box 
becomes relevant) to solutions characterized by a simple spatio-temporal 
scaling relation in which the temporal dependence of the scaling can 
be derived from the linearized fluid theory. These results are 
qualitatively very similar to those observed in numerical studies
in the context of cosmology, i.e., for expanding space-times and
for more complex initial conditions in which the displacements of 
the particles off the lattice are correlated in order to produce 
the PS of fluctuations of  cosmological models. More specifically,
the observed spatio-temporal scaling is a simple  generalization 
of what is known in the cosmology literature as  ``self-similarity'' 
in an expanding  universe 
to the case of (i) a static universe, and (ii) a PS $P(k) \propto k^2$. 
Further we
have observed that there is a transient phase to this behavior, in
which already, to a good approximation, the same spatio-temporal
scaling relation holds for the two-point correlation function
$\xi(r,t)$, but with a more rapid temporal evolution of the scaling
factor.  We have noted that the lagging of the evolution behind the
asymptotic behavior in this regime can be ascribed to effects of
discreteness (i.e. non fluid effects) slowing down the evolution of
fluctuations at scales comparable to the inter-particle distance
which have been quantified in \cite{joyce_05,marcos_06}. 
We have seen also that the form of the correlation function emerges
already at the very early times when the first non-linear correlations
develop due to two-body correlations which develop under the effect of
nearest neighbor interactions.

The gravitational evolution of a SL in a static universe thus shares
the qualitative features of similar, but more complicated models, in
cosmology. It thus provides a simplified ``toy model'' in which to
study some fundamental problems which remain open concerning the
evolution of these systems, which have been studied extensively
in numerical simulations but remain poorly understood analytically,
notably:

\begin{itemize}

\item The absence of a theory which adequately explains
the shape (i.e. functional form) and evolution of 
the observed non-linear correlations. 

\item The absence of a ``theory of discreteness errors''.
In cosmology simulations of particles displaced off lattices (or
``glasses'') aim to reproduce the evolution of a self-gravitating
fluid. There is currently very little systematic understanding of how
well this evolution is actually approximated. We have highlighted in
this paper that the SL gives a very well defined, and simplified,
framework in which to address this problem.

\end{itemize}

Let us remark finally on a few other points:

\begin{itemize}

\item We have worked here with initial velocities set equal
to zero. In exploring the analogy with cosmological simulations there
is another choice of initial velocities which is natural.  This is
that corresponding to that given by the Zeldovich approximation
discussed above, with $f(t)$ chosen in Eq.~(\ref{Zeldovich}) to
correspond to the purely growing mode of density fluctuations, 
i.e., $f(t)=e^{t/\tdyn}$.  The initial velocities are then simply the
initial displacements divided by $\tdyn$. This introduces no further
new characteristic scales in the initial conditions. Its effect on the
evolution will be to make the transient to self-similarity slightly
shorter, but it will not significantly change any of our findings or
conclusions.

\item We have made a specific choice of PDF for our shuffling,
given in Eq.~(\ref{eq:pu}). We expect different choices again to
modify slightly the nature of the transient, but not the
self-similarity. This latter, as we have emphasized, depends only on
the $k^2$ form of the PS at small $k$, 
which is in fact the same for any PDF with
finite variance. Indeed the coefficient of the $k^2$ is just given by
this variance, and the difference between PDFs will manifest
themselves in modifications of the fluctuations at small scales
(i.e. larger $k$).  For
example if the two PDF have different fourth moments, this will be
reflected in a different coefficient in the $k^4$ correction to the
small $k$ PS. Just as in the case of velocities, there is a natural
choice if one wishes to maximize the analogy with cosmological
simulations: a simple Gaussian PDF which is what is used in this
context. In fact this choice is also natural from another point of
view, as we will explain in detail in a forthcoming article\cite{baertschiger_06_prep}: 
when one
considers constructing new particle distributions by a simple
``coarse-graining'' on some scale, the SL with Gaussian PDF, due to
the Central Limit Theorem, has the property of being the unique one
which is invariant under such a coarse-graining.

\item We have reported in this paper simulations in which 
the softening $\varepsilon$ has been kept fixed (in our chosen
length units). We have mentioned that we have checked that our
results for clustering amplitudes above this scale are robust
to the use of significantly smaller values of $\varepsilon$.
A more extensive and systematic study of the role of this 
parameter would, however, be of interest, specifically
with the goal of understanding in detail how the clustering
properties are modified by it at small scales.

\end{itemize}

\acknowledgments{
We thank the ``Centro Ricerche e Studi E. Fermi'' (Roma) for the use
of a super-computer for numerical calculations, the EC grant
No. 517588 ``Statistical Physics for Cosmic Structures" and the
MIUR-PRIN05 project on ``Dynamics and thermodynamics of systems with
long range interactions" for financial support.  M.\,J. thanks the
Istituto dei Sistemi Complessi for its kind hospitality during 
October 2005 and May 2006.}. We are indebted to B. Marcos for
extensive collaboration on closely related work.


\appendix

\section{Fluid Equations and Fluid Linear Theory}
\label{app-a}

The equations which describe  the evolution of a self-gravitating
fluid are the following (e.g.
\cite[chap. II]{peebles}
or \cite[chap. 5.2]{binney})
\begin{subequations}
  \begin{align}
    &\partial_t \rho + \nabla_{\ve x}\cdot (\rho \ve v) = 0 \ , \\
    &\partial_t  \ve v  + (\ve v \cdot \nabla_{\ve x})  \ve v =
    \ve g\  - \frac{1}{\rho} \nabla_{\ve x} p\\
    & \nabla_{\ve x} \cdot \ve g = -4\pi G ( \rho -\rho_0 ) \ ,
    \label{eq:swindle} \\
    & \nabla_{\ve x} \times \ve g = 0 \ ,
  \end{align}
  \label{eq:fluid}
\end{subequations}
where $\rho(\ve x,t)$ is the mass density, $\ve v(\ve x,t)$ the
velocity field, $\ve g(\ve x,t)$ the gravitational field 
and $p(\ve x,t)$ the pressure. The set of equations closes
if $p(\ve x,t)$  is specified as a function of the density.

As it is shown in \cite{peebles,saslaw2000}, this
set of equations can be obtained after certain approximations 
from the Vlasov equation coupled to the Poisson equation:
\begin{subequations}
  \begin{equation}
    \left[ \partial_t  + \ve v\cdot \nabla_{\ve x}  - \nabla_{\ve v} 
      \Phi  \cdot
      \nabla_{\ve v} \right] f(\ve x,\ve v,t) = 0
  \end{equation}
  where $\Phi$ satisfies the modified Poisson equation
  \begin{equation}
    \nabla_{\ve x}^2 \Phi(\ve x,t) = 4\pi G \left [ m \int_{\mathbb R^3}
      f(\ve x,\ve v,\ve t)\, \D[3] \ve v -\rho_0 \right]
  \end{equation}
  \label{eq:vlasovpoisson}
\end{subequations}
and $f(\ve x,\ve v,t)$ is the density of particles in the
infinitesimal volume $\D[3]\ve x\, \D[3]\ve v$ at $(\ve x,\ve v)$ at
time $t$. These equations can themselves be derived as truncations
of a BBGKY hierarchy\cite{peebles,saslaw2000}, or starting
from a Liouville equation for the full (``spiky'') one particle
phase space density \cite{buchert_dominguez, spohn}. 

By performing a perturbation analysis ,for the case of pressureless
(i.e. highly non-relativistic or ``cold'') matter,  
around $\rho=\rho_0$ and $\ve v
= \ve 0$ with the set of Eq.~\eqref{eq:fluid}, one finds at first
order that the evolution of the density contrast $\delta(\ve
x,t)=(\rho-\rho_0)/\rho_0$ is described by the differential equation
\begin{equation}
  \ddot{\delta}(\ve x,t) = 4\pi G \rho_0 \delta(\ve x,t) 
\label{eq:ddotdelta}
\end{equation}
or equivalently that each Fourier mode $\tilde\delta(\ve k,t)$ evolves
independently of the others:
\begin{equation}
  \ddot{\tilde\delta}(\ve k,t) = 4\pi G \rho_0 \tilde\delta(\ve k,t) \ .
  \label{eq:deltakfluid}
\end{equation}

The general solution of Eq.~\eqref{eq:ddotdelta} is \cite[\S 13]{peebles}
\begin{equation}
\delta(\ve x,t ) = A(\ve x) \left( \sqrt{4\pi G \rho_0}\
  t\right) + B(\ve x) \exp \left(- \sqrt{4\pi G \rho_0}\
  t\right) \,.
 \label{eq:delta_static}
\end{equation}
For the case, as in this paper, in which the initial velocity is set 
equal to zero, one obtains
\begin{equation}
  \delta(\ve x,t) = \delta(\ve x,0) \cosh\left( \sqrt{4\pi G \rho_0}\
  t\right) \;. \label{eq:densitycontrastlinear} 
\end{equation}

\section{Lagrangian fluid theory and the Zeldovich approximation}
\label{app-b}

The previous appendix uses the Eulerian formalism of fluid mechanics,
in which one describes the evolution of the different quantities 
characterizing the fluid (velocity, density and pressure) at each point 
of a fixed reference frame. In the alternative Lagrangian formalism
(see, e.g.,\cite{buchert1,buchert2,sahni_95}), one
describes the evolution of the fluid in terms of the displacements of
its elements with respect to a reference frame. In that case, the
equations~\eqref{eq:fluid} which describes the density and the
velocity are then transformed into a set of equation describing the
evolution of a displacement field $\ve f(\ve X,t)$. The position $\ve
x$ of a fluid element at time $t$ is then written as
\begin{equation}
\ve x = \ve f(\ve X,t)  
\end{equation}
where the coordinate $\ve X$ labels the fluid element considered. One
can choose this coordinate as the position of the fluid element at
the initial time (which we assume to be 0): $\ve X = \ve f(\ve X,0)$. The
equations for $\ve f(\ve X,t)$ in the case of a gravitating fluid can
be found in, e.g., \cite{buchert2}. Note that in this reference, a
fluid in an expanding universe is considered. The static case can be
recovered by setting the expansion factor $a(t)=1$ at all 
times.

As in the Eulerian approach, one can perform a perturbation theory in
the Lagrangian approach. One writes $\ve f(\ve X,t) = \ve
X + \ve p(\ve X,t)$ with $\ve p(\ve X,0)=\ve 0$, and 
performs a Taylor  expansion in powers of $\ve p$.
At linear order in $\ve p(\ve X,t)$, one obtains the following
set of equations:
\begin{align}
  &\nabla\cdot (\ddot {\ve p} - 4\pi G \rho_0 \ve p) = - 4 \pi G
  \rho_0 \delta(\ve X,0) \\ &\nabla\times \ddot{ \ve p} = \ve 0
\end{align}
where $\delta(\ve X,0)$ is the density contrast at $t=0$.
Writing the
vector field $\ve p$ as the sum of a curl-free part $\ve p_D$ and a
divergence-less part $\ve p_R$ (i.e. $\ve p_D$ can be written
as the gradient of a scalar function, and $\ve p_R$ as the
curl of a vector field), one finds after some calculation that
\begin{equation}
\begin{split}
  \ve p(\ve X,t) = &\ddot{\ve p} (\ve X,0) \frac{\left[ \cosh\left(
	\sqrt{4\pi G \rho_0}\ t \right) -1 \right] }{4\pi G \rho_0} \\
	&+ \dot{\ve p}_D(\ve X,0) \frac{\sinh\left( \sqrt{4\pi G
	\rho_0}\ t\right)}{\sqrt{4\pi G \rho_0}} + \dot{\ve p}_R(\ve
	X,0) t
\end{split}
 \label{eq:psolution}
\end{equation}
with the initial condition $\ve p(\ve X,0)=\ve 0$. Note that
$\ddot{\ve p}(\ve X,0)=\ddot{\ve p}_D(\ve X,0)$ since the
gravitational force is conservative. These 
Eqs.~(\ref{eq:psolution}) correspond to Eqs.~(6) and ~(7)
in \cite{buchert2}, with $a(t)=1$ and $\Lambda = 4\pi G \rho_0$.

The asymptotic behavior of the solution~\eqref{eq:psolution} is
\begin{equation}
  \ve {p}(\ve X,t) \xrightarrow[t\to\infty]{} \frac{1}{2} \left[
  \frac{\ddot{\ve p} (\ve X,0)}{4\pi G \rho_0} + \frac{\dot{\ve
  p}_D(\ve X,0)}{\sqrt{4\pi G \rho_0}} \right] \exp\left( \sqrt{4\pi G
  \rho_0}\ t\right) \ .
\end{equation}
By choosing $\dot{\ve p}_R(\ve X,0)=\ve 0$ and $\ddot{\ve p}(\ve X,0)
\sqrt{4\pi G \rho_0} = \dot{\ve p}(\ve X,0)$, the solution is then
directly in its asymptotic regime. This is the static space 
equivalent of the Zeldovich approximation in an
expanding background\cite{zeldovich_70,shandarin_zeldovich,buchert2,sahni_95}.

The linear approximation of the Lagrangian approach, which leads to
the Zeldovich approximation as we have described, has proven to be
very useful in the problem of gravitational clustering. With respect
to the linear Eulerian approach, it has the advantage that it
can describe the evolution of density fluctuations with a 
density contrast much greater than unity.

\end{document}